\documentclass[aps,prd,twocolumn,floatfix, superscriptaddress]{revtex4-2}

\usepackage{amsmath}
\usepackage{amssymb}
\usepackage{amsfonts}
\usepackage{amsthm}
\usepackage{epsfig}
\usepackage{graphicx}
\usepackage{color}
\usepackage[normalem]{ulem}
\usepackage{url}
\usepackage{algorithm}
\usepackage{algpseudocode}
\usepackage{orcidlink}

\begin{document}

\title{A new framework for lightning-fast gravitational wave analysis of pulsar timing data}

\author{Aiden Gundersen\orcidlink{0009-0004-2085-6348}}
\affiliation{eXtreme Gravity Institute, Department of Physics, Montana State University, Bozeman, Montana 59717, USA}

\author{Rutger van Haasteren\orcidlink{0000-0002-6428-2620}}
\affiliation{Max Planck Institute for Gravitational Physics (Albert Einstein Institute), 30167 Hannover, Germany
Leibniz Universit\"{a}t Hannover, 30167 Hannover, Germany}

\author{Neil J. Cornish\orcidlink{0000-0002-7435-0869}}
\affiliation{eXtreme Gravity Institute, Department of Physics, Montana State University, Bozeman, Montana 59717, USA}

\author{Michele Vallisneri\orcidlink{0000-0002-4162-0033}}
\affiliation{ETH Zurich, Institute for Particle Physics and Astrophysics, Wolfgang-Pauli-Strasse 27, 8093 Zurich, Switzerland}

\author{Patrick M. Meyers\orcidlink{0000-0002-2689-0190}}
\affiliation{ETH Zurich, Institute for Particle Physics and Astrophysics, Wolfgang-Pauli-Strasse 27, 8093 Zurich, Switzerland}

\begin{abstract} 
 Pulsar timing array data analysis is computationally expensive, limiting the complexity of models which can be studied. As pulsar timing datasets and their respective models grow in size and sophistication, faster and scalable inference methods are essential. In this paper, we accelerate pulsar timing analyses by sampling in the space of Fourier coefficients instead of analytically marginalizing over them. Previous studies have shown the Fourier space induces a complex, high-dimensional posterior geometry, from which it is generally difficult to sample. We show that under an appropriate coordinate transformation the Fourier coefficients approximately follow a standard normal distribution, and may be efficiently sampled using a Hamiltonian Monte Carlo scheme. Under this coordinate transformation, for datasets of size and complexity comparable to the NANOGrav 15-year release, the new method produces converged posterior distributions for a range of models which include inter-pulsar correlations, stochastic, and deterministic signals in approximately 15 minutes on an NVIDIA GeForce RTX 3090 GPU. By comparison, the legacy pulsar timing analysis software \texttt{ENTERPRISE} would require months of computation on a CPU cluster to analyze comparable datasets under the same joint stochastic and deterministic models.
\end{abstract}

\maketitle

\section{Introduction}
Pulsar timing arrays (PTAs) measure the times of arrival (TOAs) of radio pulses from pulsars. If TOA observations span decades over a network of pulsars distributed across the sky, then PTAs are sensitive to gravitational waves (GWs) of nanohertz frequencies \citep{Sazhin1978, Detweiler}. Presently, the North American Nanohertz Observatory for Gravitational Waves (NANOGrav), the Parkes Pulsar Timing Array, the European Pulsar Timing Array in collaboration with the Indian Pulsar Timing Array, the Chinese Pulsar Timing Array, and the MeerKAT Pulsar Timing Array have found evidence for a stochastic gravitational wave background (GWB), \citep{NG15, ParkesPTA, EPTA, ChinesePTA, MeerKAT_PTA}.

There are many hypotheses for what physical process gives rise to the stochastic GWB, the most popular being the background is realized by an incoherent superposition of GWs emitted by a population of supermassive black hole binaries (SMBHBs), formed during galaxy mergers \citep{GWB1, GWB4, GWB3, GWB5, GWB6, GWB7}. The background may also include contributions from cosmological phase transitions, primordial GWs, and other exotic sources, see e.g. \cite{NG15_new_physics} and references therein.

Besides the GWB, other signal and noise processes, covariant with the background, populate PTA datasets and must be jointly modeled. This includes, but is not limited to, radiometer noise in the telescopes, achromatic red noise intrinsic to the pulsars, stochastic fluctuations of the dispersion of radio pulses in the interstellar medium, and GWs from individual SMBHBs which are particularly massive or nearby, and discernible from the background population.

Standard analyses of PTA datasets use numerical Bayesian inference, where a (time-domain) posterior probability density is constructed. The posterior describes the probability of model parameters conditioned on the data observed. Inference is performed by sampling the posterior with Markov Chain Monte Carlo (MCMC) methods, requiring many evaluations of the posterior. Recently, other approaches such as stochastic gradient-descent Bayesian variational inference and simulation based inference with GPU implementations have accelerated the analysis \citep{PTA_VI, SBI_in_PTAs}.

PTA data analysis is computationally expensive. Typical datasets consist of hundreds of thousands of radio pulse TOAs recorded over decades. The observations are unevenly spaced and the noise heteroscedastic, necessitating the analysis to be conducted in the time-domain. Moreover, radio pulse timing delays due to pulsar proper motion, spin period, and other deterministic effects are fit using a per-pulsar time-domain model. The substantial number of TOAs in realistic datasets requires large and expensive linear algebra computations on the order of the size of the dataset. The analysis problem only worsens as datasets grow with more observations and as pulsars are added to the array. Complicated models further bottleneck the analysis as they generally involve higher-dimensional parameters spaces and expensive posterior evaluations which require more  calls to adequately sample the distribution. Currently, the complexity of models used in PTA analyses are restricted by their computational burden, rather than the resolution of the data.

The aim of this paper is to accelerate Bayesian analyses of GWs in PTA datasets, allowing more in-depth modeling of larger datasets. The latest PTA analysis packages have leveraged the \texttt{JAX} \cite{JAX} software package, GPU-acceleration, and gradient-based samplers to outpace the flagship software \texttt{ENTERPRISE} \cite{Enterprise}, but continue to analytically marginalize over a large number of Fourier coefficients. The methods presented in this paper can analyze the latest publicly released datasets over an order of magnitude faster than even these accelerated approaches by sampling in the space of Fourier coefficients (i.e. by performing the marginalization numerically rather than analytically). We reproduce some results from the NANOGrav 15-year stochastic analysis \citep{NG15} with these methods and perform previously computationally inaccessible analyses on simulated datasets. These speedups are achieved by applying a coordinate transform on the hyper-efficient posterior presented in \cite{Lentati}. This posterior, while hyper-efficient, is high-dimensional and exhibits a complex funnel geometry from which it is difficult to sample. The coordinate transform reparameterizes the funnel so the posterior more closely resembles a standard normal distribution, from which samples may be drawn very efficiently. The inverse transformation maps samples back to those of the original, untransformed, posterior.

Previous works have sampled this hyper-efficient, but geometrically difficult posterior to varying degrees of success. \cite{Lentati} uses a Hamiltonian Monte Carlo (HMC) sampling algorithm to sample the posterior distribution in the high signal-to-noise regime where the funnel did not manifest, \cite{vH_Gaussian_advances,PTA_Gibbs23, PTA_Gibbs25} uses Gibbs sampling on a subset of models with a weak-enough funnel to sample efficiently, and \cite{joint_CW, escaping_funnel} extends the posterior to include deterministic signals, and uses a generalized (even higher-dimensional) model parameterized to weaken the sharpness of the funnel. Other approaches have applied coordinate transformations to PTA analyses, e.g. \cite{Free_spec_whitened} sampled in the space of whitened power per frequency bin, our approach on the other hand applies a coordinate transformation to the latent mode amplitudes per frequency bin which directly represent the data.

Primarily, previous studies have focused on mitigation and sampling techniques \textit{after} the construction of the posterior. Our objective, based on the work of \cite{vallisneriTamingOutliersPulsartiming2017}, is to reparameterize the posterior from the start. That way, increasingly intricate sampling techniques need not be introduced to retain efficiency as the complexity of models and the size of datasets grow in the future. While the original approach of \cite{vallisneriTamingOutliersPulsartiming2017} is computationally similar to our transformation of the posterior distribution, it was so extremely inefficiently implemented that it was computationally challenging even for single-pulsar analyses. Our current implementation is many orders of magnitude faster due to our Fourier-domain focus, parallelization, and the use of \texttt{JAX} \citep{JAX}. Moreover, our approach generalizes the coordinate transformation to include contributions from deterministic signals, inter-frequency correlations, and non-Gaussian features.

Despite its near normal distribution, the transformed posterior remains difficult to sample due to its high-dimension, having $\mathcal{O}(10^3)$ parameters (under commonly used models for the latest datasets). For this reason we sample with HMC and a No U-Turn Sampler (NUTS) scheme, which is inspired by the Hamiltonian structure of classical mechanics (see e.g. \cite{concept_HMC}). HMC integrates trajectories through a phase space, in which the target posterior is embedded, sampling along the way. The integrated trajectories are deterministic, allowing the sampler to traverse larger distances through parameter space than alternative random walk proposals. It is advantageous for high-dimensional densities where, under careful tuning, it can achieve high (often $\sim90\%$) proposal acceptance rates and low auto-correlation lengths. For example, the computational cost to sample a $d$-dimensional multivariate normal distribution with naive random-walk Metropolis is $\mathcal{O}(d^2)$, \cite{RWM_scaling}, while the cost with HMC is $\mathcal{O}(d^{5/4})$ \cite{HMC_scaling}. Such sampling schemes have previously been used in PTA data analysis, \cite{PTA_HMC1, PTA_HMC2}, but the posterior sampled was one in which nuisance parameters were analytically marginalized from the model rendering a lower-dimensional, but computationally expensive, target distribution bottlenecking the sampler. The HMC sampling scheme is summarized in Appendix~\ref{app:HMC}.

Altogether, the methods presented here allow us to analyze large PTA datasets under joint (stochastic and deterministic) models over an order of magnitude more efficiently than previous approaches. Section~\ref{sec:HBM} reviews Bayesian hierarchical modeling and demonstrates the effectiveness of standardizing coordinate transformations on a toy example. The signal and noise components of PTA analyses, and their respective models, are discussed in Sec.~\ref{sec:signal_comps}. Section~\ref{sec:HBM_PTA} constructs the PTA hierarchical Bayesian model and the coordinate transformation under which the posterior may be efficiently sampled. The coordinate transform is first presented for a purely stochastic model, then generalized to include deterministic contributions and inter-frequency correlations. Previous results from the NANOGrav 15-year analysis \citep{NG15} are reproduced Sec.~\ref{sec:Results} along with an analysis of simulated data.

\section{Inference in Hierarchical Models}\label{sec:HBM}
Hierarchical models are powerful tools in Bayesian inference. Generally a set of low-level (or latent) parameters, $\mathbf{x}$, are used to describe the data, $\mathbf{d}$, for which a likelihood function may be constructed, $p(\mathbf{d}|\mathbf{x})$. Additional (high-level) hyper-parameters, $\mathbf{y}$, are used to parameterize the prior distribution of the low-level parameters, $p(\mathbf{x}|\mathbf{y})$, and a hyper-prior is placed on the hyper-parameters, $p(\mathbf{y})$. Bayes' theorem allows us to construct the full hierarchical model,
\begin{equation}\label{eq:gen_HBM}
    p(\mathbf{x}, \mathbf{y}|\mathbf{d})\propto p(\mathbf{d}|\mathbf{x})\cdot p(\mathbf{x}|\mathbf{y})\cdot p(\mathbf{y})\,.
\end{equation}

Such hierarchies are common when modeling data of individuals, while simultaneously wishing to conduct inference on the population. For example, the LIGO and VIRGO detectors, \citep{LIGO, VIRGO}, have observed hundreds of short duration GWs, \citep{GWTC5}. Hierarchical models are necessary when inference is desired simultaneously on the parameters of individual compact binaries as well as on the parameters of the population. The parameters for each individual binary are the low-level parameters, whose priors are conditioned on the population parameters \cite{Adams:2012qw,BHpop1,BHpop2, BHpopHierarch, Taylor_hierarchical}. Beyond being useful tools, hierarchical models are required for the analysis of PTA datasets \citep{RvH_PTAs_require_HBM, Boris_hierarchPTA, Boris_hierarchPTA2}. For a more in-depth discussion of Bayesian hierarchical models, see e.g. Ch. 5 of \cite{BDA_Gelman}.

\subsection{Neal's funnel}
While extremely powerful, the coupling between low- and high-level parameters induced by hierarchical models can yield complicated posterior geometries, from which it is difficult to sample with standard techniques. Neal's funnel is one such geometry, referring to an exponentially tapering probability density \citep{NealsFunnel}. The funnel is present in regions of parameter space where a hyper-parameter dictates the variance of a low-level parameter. The opening of the funnel is formed where the low-level parameter is allowed significant variance and the throat where the variance is constricted. Standard MCMC methods, such as Random Walk Metropolis \citep{Metropolis}, struggle to sample distributions with such geometry because increasingly precise jump proposals must be made as the chain traverses the throat of the funnel. In practice, naive samplers will get stuck in the throat of the funnel, and fail to explore the target distribution.

There are several ways to improve sampling for densities exhibiting Neal's funnel. More robust samplers, such as Riemannian Manifold Hamiltonian Monte Carlo, have been shown to effectively sample funnels \citep{RMHMC1, RMHMC2}. In some cases, the latent parameters of the funnel may be analytically marginalized from the hierarchical model. Analytic marginalization results in a lower-dimensional, funnel-less, and overall easier parameter space to explore. However, the marginalized target density is often significantly more computationally expensive to evaluate, bottlenecking the sampler, as is the case of standard PTA analyses. The funnel may also be avoided by generalizing the parameter space to a higher-dimensional space in which the sharpness of the funnel is lessened. Once sampled, the higher-dimensional distribution may be constrained to lie in the original space containing the funnel \citep{escaping_funnel, joint_CW}. Lastly, one may perform a coordinate transformation, effectively reparameterizing the model so its geometry is more easily explored \citep{ReparamFunnel}. In this paper, we will address the funnel geometry of the PTA posterior with such a coordinate transformation.

\subsection{Reparameterization and the standardizing transform}\label{subsec:gen_reparam}
Consider a set of random variables $\mathbf{x}$ which are distributed according to some probability density function, $p = p(\mathbf{x})$. The density may be arbitrarily reparameterized under a bijective and differentiable coordinate transformation, $T$, via the mapping $\mathbf{z} = T(\mathbf{x})$ and its inverse $\mathbf{x} = T^{-1}(\mathbf{z})$. The transformed density, $\tilde{p}$, is defined over the new coordinates, $\tilde{p}=\tilde{p}(\mathbf{z})$, and must conserve probability mass, $p\,d\mathbf{x} = \tilde{p}\,d\mathbf{z}$ from which it follows
\begin{equation}\label{eq:coor_transform}
    \tilde{p}(
    \mathbf{z}) = p(\mathbf{x})\cdot\text{det}(\partial\mathbf{x}/\partial\mathbf{z}) = p\big(T^{-1}(\mathbf{z})\big)\cdot\text{det}(\partial\mathbf{x}/\partial\mathbf{z})
\end{equation}
where $\text{det}(\partial\mathbf{x}/\partial\mathbf{z})$ denotes the determinant of the Jacobian of the coordinate transformation. If we wish to draw random samples $\mathbf{x}$ from the target distribution, we may either sample the target distribution, $p$, directly or sample in $\mathbf{z}$ from the transformed distribution, $\tilde{p}$, and map our samples back to the original coordinates under the inverse transformation, $\mathbf{x} = T^{-1}(\mathbf{z})$. Both methods produce identical distributions of the random variables, $\mathbf{x}$, provided the coordinate transformation is bijective, differentiable, and has a non-vanishing Jacobian determinant.

While any well-behaved coordinate transformations may be employed, the most effective is one which yields a transformed target distribution from which it is easiest to sample. Generally the more a target density resembles an uncorrelated standard normal distribution, the more efficiently independent samples may be drawn. An effective reparameterization is then one in which the target density is ``standardized". We will refer to such transformations as \textit{standardizing transforms} (also known as non-centered parameterizations, decentering transforms, or whitening transforms); see \cite{ReparamFunnel} for a rigorous discussion of such transformations in Bayesian hierarchical models.

To perform a standardizing transformation, the mean and covariance of the target density, $p(\mathbf{x})$, are estimated, $\boldsymbol{\mu} = \mathbb{E}[\mathbf{x}]$ and $\boldsymbol{\Sigma} = \mathbb{E}[(\mathbf{x}-\boldsymbol{\mu})(\mathbf{x}-\boldsymbol{\mu})^\text{T}]\equiv\text{cov}(\mathbf{x}, \mathbf{x})$, and the standardizing transformation and its inverse are
\begin{equation}\label{eq:standardizing_transform}
    \mathbf{z} = T(\mathbf{x}) = \mathbf{L}^{-1}(\mathbf{x} - \boldsymbol{\mu}) \;\leftrightarrow \; \mathbf{x} = T^{-1}(\mathbf{z}) = \boldsymbol{\mu} + \mathbf{L}\mathbf{z}\;
\end{equation}
where $\mathbf{z}$ are the new \textit{standardized} (or rather \textit{whitened}) coordinates and $\mathbf{L}$ is the Cholesky decomposition of the covariance matrix, $\boldsymbol{\Sigma} = \mathbf{L}\mathbf{L}^\text{T}$. It is straightforward to check if $p(\mathbf{x})$ is well-approximated by the Gaussian moments, $\boldsymbol{\mu}$ and $\boldsymbol{\Sigma}$, then $\tilde{p}(\mathbf{z})$ is approximately an uncorrelated standard normal density. Substituting Eq.~(\ref{eq:standardizing_transform}) into Eq.~(\ref{eq:coor_transform}), the relationship between the original target density and its standardized form is
\begin{equation}\label{eq:standardized_density}
    \tilde{p}(\mathbf{z}) = p(\boldsymbol{\mu} + \mathbf{L}\mathbf{z})\cdot\text{det}(\mathbf{L})\,.
\end{equation}
After samples are obtained from the standardized density Eq.~(\ref{eq:standardized_density}), random samples from the original target distribution may be calculated using the coordinate transformation, Eq.~(\ref{eq:standardizing_transform}). It is generally very efficient to sample from near-normal distributions like Eq.~(\ref{eq:standardized_density}) with standard sampling schemes. Note that the statistical moments, $\boldsymbol{\mu}$ and $\boldsymbol{\Sigma}$, need not be exact, and the target density may exhibit significant higher-order statistical moments, beyond covariance (see Appendix \ref{app:non-gauss}). We may still sample from the true target density with Eq.~(\ref{eq:standardized_density}), but the standardizing transformation may lose its effectiveness, converging slower, as the Gaussian approximation weakens.

\subsection{Sampling a toy model}\label{sec:toy_funnel}
In this section we define a simple ``toy" hierarchical Bayesian model with a funnel, similar to that first explored in \cite{NealsFunnel}. We sample the toy model both directly and under a standardizing transform to demonstrate the effectiveness of reparameterization. The toy hierarchical model is,
\begin{equation}\label{eq:toy_HBM}
    p(\mathbf{x}, y) = p(\mathbf{x}|y)\cdot p(y),
\end{equation}
with support for low-level parameters $\mathbf{x}\in\mathbb{R}^d$ for some integer $d \geq 1$ and a hyper-parameter $y\in\mathbb{R}$. The conditional density and hyper-prior are
\begin{align}\label{eq:toy_dist}
    &\mathbf{x}|y\sim\mathcal{N}(y^2\,\mathbf{1},e^{y}\,\mathbf{I})\nonumber \\
    &y\sim\mathcal{N}(0, 9),
\end{align}
where $\mathbf{1}$ is the $d$-dimensional vector of ones, $\mathbf{I}$ is the identity matrix, and $\mathcal{N}(\boldsymbol{\mu}, \boldsymbol{\Sigma})$ denotes a multivariate normal distribution with mean $\boldsymbol{\mu}$ and covariance $\boldsymbol{\Sigma}$.

Neal's funnel is observed in the low-level parameters, $\mathbf{x}$. When the hyper-parameter $y$ is negative (positive) the covariance of the low-level parameters is small (large), and the throat (opening) of the funnel is formed. Naive samplers will struggle to explore regions of the parameter space where $y$ is sufficiently negative. We therefore seek to apply a standardizing transform to the low-level parameters. The mean and covariance of the low-level parameters are $\mathbb{E}[\mathbf{x}] = y^2\,\mathbf{1}$ and $\text{cov}(\mathbf{x},\mathbf{x}) = e^y\,\mathbf{I}$, respectively. Note the mean and covariance are themselves parameterized by the hyper-parameter, $y$, and the standardizing transform is similarly parameterized,
\begin{equation}\label{eq:toy_decenter}
    (\mathbf{x},\; y) = T^{-1}(\mathbf{z}, y) = (y^2\,\mathbf{1} + e^{y/2}\,\mathbf{z}, \;y)
\end{equation}
where $\mathbf{z}$ are the standardized coordinates. As presented here, the standardizing transformation only whitens the low-level parameters, $\mathbf{x}$, and applies the identity mapping to the hyper-parameter, $y$. The determinant of the Jacobian of the transformation is $\text{det}(\partial (\mathbf{x}, y)/\partial(\mathbf{z}, y)) = e^{y\cdot d/2}$. The standardized probability density is then
\begin{equation}\label{eq:toy_standardized_density}
    \tilde{p}(\mathbf{z}, y) = p(y^2\,\mathbf{1} + e^{y/2}\,\mathbf{z}, \;y)\,\cdot\,e^{y\cdot d/2}\,.
\end{equation}
After sampling from the standardized density, Eq.~(\ref{eq:toy_standardized_density}), in which Neal's funnel should be absent, samples from the original target density, $p(\mathbf{x}, y)$, are obtained using Eq.~(\ref{eq:toy_decenter}).

\begin{figure}
    \centering
    \includegraphics[width=0.95\linewidth]{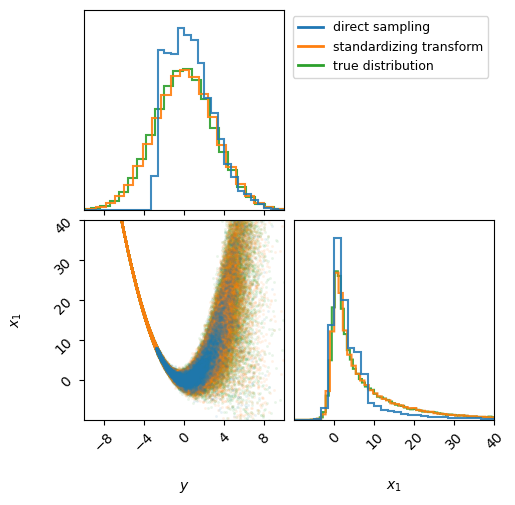}
    \caption{Samples from the hierarchical toy model, Eq.~(\ref{eq:toy_HBM}) and Eq.~(\ref{eq:toy_dist}). The sampling is conducted for $d=8$, but only samples for the hyper-parameter $y$ and the first low-level parameter $x_1$ are shown in the figure. The blue distribution attempts to directly sample the target density, and is unable to effectively explore the throat of the funnel. The orange distribution samples under a standardizing transform and is able to move through constricted regions of parameter space efficiently. The green distribution is independent draws from the target density. The samples obtained through the standardizing transform are consistent with those from the true distribution.}
    \label{fig:toy_model}
\end{figure}

We sample the toy hierarchical Bayesian model for $d=8$ both directly, and under a standardizing transformation using HMC with a NUTS scheme. Samples from both methods are shown in Fig.~\ref{fig:toy_model}. The chain attempting direct sampling of the target distribution gets stuck in the throat of the funnel and is unable to resolve the distribution. The chain sampling under the standardizing transform is able to efficiently explore the parameter space, including constricted regions in the throat of the funnel, and obtains samples consistent with the target distribution.

\section{Signal and noise components of pulsar timing analysis}\label{sec:signal_comps}
The radio pulse TOAs for a pulsar consist of deterministic and stochastic contributions,
\begin{equation}
    \mathbf{t}_\text{TOA} = \mathbf{t}_\text{det} + \mathbf{t}_\text{stoch}
\end{equation}
where $\mathbf{t}_\text{TOA}$ are the $n$ observed TOAs, $\mathbf{t}_\text{det}$ are the deterministic components, and $\mathbf{t}_\text{stoch}$ stochastic. The deterministic part of the TOAs is described primarily by the timing model, constructed per-pulsar, which models pulsar spin period, spin derivative, proper motion, and other deterministic effects \citep{NG_timing}. In the main body of this paper, the stochastic component of the TOAs is modeled as Gaussian random processes. Non-Gaussian features are discussed in Appendix~\ref{app:non-gauss}.

The timing model depends on $m$ parameters, $\boldsymbol{\beta}$. The per-pulsar timing analysis yields a set of reference timing model parameters, $\boldsymbol{\beta}_0$, used to construct the timing residuals,
\begin{equation}
    \boldsymbol{\delta t} = \mathbf{t}_\text{TOA} - \mathbf{f}(\boldsymbol{\beta}_0),
\end{equation}
where $\boldsymbol{\delta t}$ are the timing residuals and $\mathbf{f}$ the timing model. The timing residuals are expanded in terms of other signal and noise components,
\begin{align}\label{eq:residual components}
    \boldsymbol{\delta t} = \boldsymbol{\delta t}_\text{TM} &+ \boldsymbol{\delta t}_\text{WN} + \boldsymbol{\delta t}_\text{RN} + \boldsymbol{\delta t}_\text{DM} + \nonumber \\ 
    &\boldsymbol{\delta t}_\text{GWB} + \boldsymbol{\delta t}_\text{det} + \dots
\end{align}
where the various delays, $\boldsymbol{\delta t}_i$, correspond to small deviations in the timing model, white noise contributions, intrinsic instabilities in the pulsars, dispersion in the interstellar medium, a gravitational wave background, and deterministic delays not included in the timing model. The models for each of these components is summarized below.

\subsection{Linearized timing model}
We assume the set of timing model parameters which precisely describe the proper motion and spin of the pulsar are near the reference parameters. The timing model is then expanded to linear order, centered at the reference parameters, $\boldsymbol{\beta}_0$,
\begin{equation}\label{eq:TM=Me}
    \boldsymbol{\delta t}_\text{TM} = \mathbf{M}\boldsymbol{\epsilon}
\end{equation}
where $\mathbf{M}$ is the $(n\times m)$ timing design matrix with elements $M_{ij} = (\partial \mathbf{f}_i(\boldsymbol{\beta})/\partial\boldsymbol{\beta}_j)\vert_{\boldsymbol{\beta}_0}$. The parameter vector $\boldsymbol{\epsilon}=\boldsymbol{\beta}-\boldsymbol{\beta_0}$ is the linear deviation from the reference parameters. Rather than modeling the parameters of the deterministic timing model themselves, the degrees of freedom associated to the timing model are represented by the linear deviations from the reference point.

\subsection{White noise}
A TOA is constructed by de-dispersing and folding pulses observed within an observing epoch, fitting an average pulse template, and assigning an uncertainty to the TOA measurement, which is due primarily to radiometer noise in the telescope. Not all template fitting uncertainties may be propagated into the final quoted uncertainty for the TOA, so an \textit{extra factor} (EFAC) is introduced, which multiplicatively corrects uncertainties in telescope receivers and backends. Additional instrumental effects cannot be modeled by EFAC, so an \textit{extra quadrature} (EQUAD) term is added to the white noise model. The phenomenological white noise model is represented with the covariance matrix,
\begin{equation}\label{eq:WN_cov}
    \mathbf{N} = \mathbb{E}[\boldsymbol{\delta t}_{\text{WN}, i\mu}\,\boldsymbol{\delta t}_{\text{WN},j\nu}^\text{T}] = \mathcal{F}_\mu^2\sigma_i^2\delta_{ij}\delta_{\mu\nu} + Q_\mu^2\delta_{ij}\delta_{\mu\nu}
\end{equation}
where $\mathcal{F}$ is the EFAC, $\mathcal{Q}$ EQUAD, and $\sigma$ TOA uncertainty. Latin indices label TOA observations, $i,j\in\{1,2,\dots,n\}$ and Greek indices ($\mu,\nu,\dots$) denote specific receiver-backend systems. Lastly, folding a finite number of pulses within an observing epoch leads to pulse phase jitter. This induces an \textit{extra correlation} (ECORR) which correlates different bands, but is uncorrelated between observing epochs \cite{NG15_noise}.

EFAC, EQUAD, and ECORR values may be estimated from single-pulsar analyses before the multi-pulsar analysis is conducted. While it is possible to infer the white noise parameters simultaneously with other parameters of the multi-pulsar model, we will assume the initial white noise estimates are accurate, and fix the white noise model throughout our analysis. That is, the white noise covariance matrix (Eq.~(\ref{eq:WN_cov}) including ECORR contributions) is computed once at the beginning of our analysis from a set of independent single-pulsar analyses and held constant thereafter.

\subsection{Red and chromatic noise}
While largely stable rotators, millisecond pulsars exhibit quasi-random walk behavior in their pulse phase, period, and spindown rate due to internal instabilities \citep{PulsarIRN}. The resulting stochastic time-correlated delays are known as intrinsic pulsar achromatic red noise (RN) and denoted $\boldsymbol{\delta t}_\text{RN}$. The gravitational wave background also gives rise to a stochastic achromatic red signal in the TOAs, $\boldsymbol{\delta t}_\text{GWB}$, \citep{GWB1, GWB2, GWB3, GWB4}. These delays are modeled with a set of discrete Fourier modes,
\begin{align}\label{eq:Fourier_series}
    (\boldsymbol{\delta t}_\text{RN} + \boldsymbol{\delta t}_\text{GWB})_i &= \sum_{k=1}^{N_f}\big[a_k\sin(2\pi k\,\mathbf{t}_{\text{TOA}, i}/T) + \nonumber \\
    &\hspace{12mm}b_k\cos(2\pi k\,\mathbf{t}_{\text{TOA}, i}/T)\big]
\end{align}
where $i\in\{1,2,\dots,n\}$ indexes the TOA observation, $T$ is the the total observation span over all pulsars in the array, $k\in\{1,2,\dots,N_f\}$ indexes the frequency bin, and $N_f$ is the number of frequency bins modeled. The zero-frequency term is a component of the per-pulsar timing model, so $k=0$ is not included in the red noise model. The $(n\times N_f)$ Fourier design matrix consists of alternating columns of basis functions,
\begin{equation}
    F_{ik} = \big\{\sin(2\pi k\,\mathbf{t}_{\text{TOA}, i} / T)\,, \;\; \cos(2\pi k\,\mathbf{t}_{\text{TOA}, i} / T)\big\}\,,
\end{equation}
and compactly represents the Fourier series when multiplying the vector of Fourier coefficients $\mathbf{a}=[a_1, b_1, a_2, b_2,\dots,a_{N_f}, b_{N_f}]^\text{T}$. That is,
\begin{equation}\label{eq:RN=Fa}
    \boldsymbol{\delta t}_\text{RN} + \boldsymbol{\delta t}_\text{GWB} =  \mathbf{F}\mathbf{a}\,.
\end{equation}
In some cases it is advantageous to use distinct sets of Fourier coefficients to represent the GWB and intrinsic pulsar RN, $\mathbf{a}_\text{GWB}$ and $\mathbf{a}_\text{RN}$, respectively. The methods presented in this paper are generalized to such cases in Appendix~\ref{app:split_coeffs}.

As the radio pulses travel from the pulsar to the Earth, they propagate through the turbulent interstellar medium, and undergo frequency-dependent dispersion \citep{DM2, DM1}. Stochastic fluctuations in the dispersion measure (DM) induces chromatic red noise in the timing residuals which is also modeled by a Fourier series,
\begin{equation}
    \boldsymbol{\delta t}_\text{DM} = \mathbf{F}_\text{DM}\,\mathbf{a}_\text{DM}\,,
\end{equation}
where $\mathbf{F}_\text{DM} = \mathbf{F}/(K\nu^2_{\text{obs},i})$, $K=2.41\times 10^{-16}\;\text{Hz}^{-2}\text{cm}^{-3}\text{pc}\,\text{s}^{-1}$ is the dispersion constant, and $\nu_{\text{obs},i}$ is the radio frequency of the $i^\text{th}$ observation. We will neglect DM contributions throughout the rest of this paper for brevity. However, it is straightforward to add DM delays to the model. The Fourier series for DM should be appended to that of RN when desired:  $\mathbf{F}\mathbf{a}\rightarrow(\mathbf{F}\mathbf{a}, \mathbf{F}_\text{DM}\,\mathbf{a}_\text{DM})$.

\subsection{Deterministic signals and continuous waves}
Besides those of the timing model, there are potentially other deterministic signals in PTA datasets. For example, if an individual SMBHB is particularly nearby or massive, it may be discernible from the background population. The binary radiates a near monochromatic continuous gravitational wave (CW) and is modeled deterministically \citep{CWsPTA, CWsPTA2}. Other potential deterministic signals are GW bursts \citep{bursts1, bursts2}, perturbations to the solar system ephemeris \citep{SS_ephem, BayesEphem}, and the timing model itself. We'll denote an arbitrary deterministic signal model as $\mathbf{h}$ with parameters $\boldsymbol{\theta}$. The deterministic signal's contribution to the timing residuals is
\begin{equation}
    \boldsymbol{\delta t}_\text{det} = \mathbf{h}(\boldsymbol{\theta})\,.
\end{equation}

In Sec.~\ref{sec:Results}, we analyze a simulated dataset containing timing delays induced by an individual SMBHB to demonstrate the deterministic signal modeling methods presented in this paper. The delays are simulated consistent with the CW model presented in \citet{CWsPTA, EllisCW, EllisCWs2, Sample_CW_phase} and depends on $8 + 2N_p$ parameters, where $N_p$ is the number of pulsars in the array. 8 parameters are those of the binary itself: $\{\mathcal{M}, f_\text{CW}, \iota, \psi, h, \theta, \phi, \Phi_0\}$ which correspond to chirp mass, frequency, inclination angle, polarization, characteristic strain, sky location, and a reference phase, respectively. The other $2N_p$ parameters are the pulsar distances, $L_I$, and the phase of the CW at every pulsar, $\Phi_I$, where $I\in\{1, 2,\dots, N_p\}$ indexes the pulsars in the array. Note that the phase at each pulsar, $\Phi_I$, can be determined from the 8 binary and $N_p$ pulsar distance parameters. However, they are treated as independent parameters to smooth out the posterior geometry for ease of sampling \citep{Sample_CW_phase, CWsPTA}.

\section{Hierarchical inference in pulsar timing arrays}\label{sec:HBM_PTA}
Many parameters of PTA analyses are not modeled hierarchically, i.e. they use static priors. However, the Fourier coefficients which describe the stochastic GWB, intrinsic pulsar noise, and dispersion in the interstellar medium do use a spectral hyper-model, with a set of hyper-parameters $\boldsymbol{\eta}$. We will generalize the multi-pulsar model to include deterministic signals in Sec.~\ref{subsec:gen2det}, but will neglect such contributions for now. The hierarchical Bayesian model for PTAs is then
\begin{equation}\label{eq:PTA_HBM}
    p(\boldsymbol{\epsilon}, \mathbf{a}, \boldsymbol{\eta}|\boldsymbol{\delta t}) \propto p(\boldsymbol{\delta t}|\boldsymbol{\epsilon}, \mathbf{a})\cdot p(\mathbf{a}|\boldsymbol{\eta})\cdot p(\boldsymbol{\epsilon})\,.
\end{equation}
Eq.~(\ref{eq:residual components}) may be rearranged to yield a realization of white noise, $\boldsymbol{\delta t}_\text{WN} = \boldsymbol{\delta t} - \boldsymbol{\delta t}_\text{TM} - \boldsymbol{\delta t}_\text{RN} - \boldsymbol{\delta t}_\text{GWB}$, and the white noise covariance matrix may be computed from initial estimates of EFAC, EQUAD, and ECORR parameter values. The Gaussian likelihood is
\begin{widetext}
\begin{equation}\label{eq:likelihood}
    p(\boldsymbol{\delta t}|\boldsymbol{\epsilon}, \mathbf{a}) = \frac{1}{\sqrt{\text{det}(2\pi\mathbf{N}})}\,\text{exp}\bigg[-\frac{1}{2}\big(\boldsymbol{\delta t} - \mathbf{M}\boldsymbol{\epsilon}-\mathbf{F}\mathbf{a}\big)^\text{T}\,\mathbf{N}^{-1}\big(\boldsymbol{\delta t} - \mathbf{M}\boldsymbol{\epsilon}-\mathbf{F}\mathbf{a}\big)\bigg]\,,
\end{equation}
\end{widetext}
using Eq.~(\ref{eq:TM=Me}) and Eq.~(\ref{eq:RN=Fa}). The likelihood is factorized per-pulsar, each pulsar being endowed with unique timing residuals and constituent models. The likelihood for the full PTA is the product of the individual pulsar likelihoods. Nonetheless, we will use Eq.~(\ref{eq:likelihood}) to represent the likelihood for the entire PTA, understanding that the timing residuals and models are concatenated across pulsars.

\subsection{Priors}
The marginalized distribution of linear deviations to the timing model is dominated by the likelihood, so it is difficult to differentiate between a wide- and infinitely-wide prior. It is convention to model the linear deviations with a normal prior of zero mean and infinite variance, $\boldsymbol{\epsilon}_i\sim\mathcal{N}(0, \infty)$, where $i\in\{1,2,\dots, m\}$ indexes the parameters of the timing model. The prior on Fourier coefficients is a multivariate normal distribution of zero mean and covariance
\begin{equation}\label{eq:prior_cov}
    \boldsymbol{\phi}_{IJ\,ij} = \alpha_{IJ}\rho_i\delta_{ij} + \delta_{IJ}\kappa_{Ii}\delta_{ij}
\end{equation}
where $I, J$ label pulsars in the array, $i, j$ label the frequency bin, and $\boldsymbol{\rho}$ and $\boldsymbol{\kappa}$ denote the power spectrum of the GWB and RN, respectively. There is no summation over repeated indices. The intrinsic pulsar red noise is uncorrelated between pulsars while the GWB obeys the Hellings-Downs (HD) correlation pattern \cite{HDcorr}
\begin{equation}\label{eq:HD}
    \alpha_{IJ} = \frac{3}{2}\beta_{IJ}\ln\beta_{IJ} - \frac{1}{4}\beta_{IJ} + \frac{1}{2} + \frac{1}{2}\delta_{IJ}
\end{equation}
where $\beta_{IJ}=(1-\cos\Theta_{IJ})/2$ and $\Theta_{IJ}$ is the angle between pulsars $I$ and $J$ on the sky.

Arbitrary spectral models may be used to describe the intrinsic pulsar RN and stochastic GWB. One common choice is a free spectral model, where the power is allowed to vary freely in each frequency bin. That is, each $\rho_i$ (or $\kappa_{Ii}$) is itself a free (hyper-) parameter. Another choice is a power law spectral model, which parameterizes the spectrum with an amplitude and spectral index,
\begin{equation}\label{eq:power_law}
    \rho_i(A, \gamma) = \frac{A^2}{12\pi^2}\frac{1}{T}\left(\frac{i/T}{1\,\text{yr}^{-1}}\right)^{-\gamma}\,\text{yr}^2\,,
\end{equation}
where $A$ and $\gamma$ are the amplitude and spectral index of the power spectrum, respectively. We choose a reference frequency $f_\text{ref} = \text{yr}^{-1}$. The power law may model the GWB common to all pulsars $(A_\text{GWB}, \gamma_\text{GWB})$, or intrinsic RN $(A_I, \gamma_I)$ unique to each pulsar. Arbitrary spectral models are amenable to the methods presented in this paper and are in use throughout PTA analyses, see e.g. \cite{turnover, spectrum_knees}. The covariance matrix may even be parameterized by galaxy stellar mass functions, merger rates, and other astrophysical parameters from which the GWB spectrum is derived \cite{NG15_new_physics, Holodeck}. Let $\boldsymbol{\eta}$ denote the hyper-parameters for an arbitrary hyper-model, describing the spectrum of both the stochastic GWB and pulsar RN. The prior on the Fourier coefficients is
\begin{equation}\label{eq:a_prior}
    p(\mathbf{a}|\boldsymbol{\eta}) = \frac{1}{\sqrt{\text{det}(2\pi\boldsymbol{\phi})}}\,\text{exp}\bigg[-\frac{1}{2}\mathbf{a}^\text{T}\,\boldsymbol{\phi}^{-1}\,\mathbf{a}\bigg],
\end{equation}
where the covariance is determined from a set of hyper-parameters and choice of spectral model, $\boldsymbol{\phi} = \boldsymbol{\phi}(\boldsymbol{\eta})$.

Historically the power law, Eq.~(\ref{eq:power_law}), was favored for both intrinsic pulsar RN and GWB models in many PTA analyses. It was advantageous requiring only two hyper-parameters (per pulsar and GWB), keeping the dimensionality of the posterior relatively low. Moreover, general relativity predicts $\gamma_\text{GWB} = 13/3$ if the background is realized via a population of circularly inspiraling binaries \cite{GWB1, GWB2, GWB3}. While the free spectral model is more flexible, it was previously restricted in applications due to the computational cost of its dimension (most analyses could only afford to model the GWB and a small subset of pulsars with a free spectrum). By assuming statistical independence across frequency bins, despite uneven sampling of the data in the time-domain inducing such correlations \cite{Lentati, vH_Gaussian_advances, TD_free_spec}, factorized likelihood methods \citep{factorized_like1, factorized_like2} are able to efficiently perform free spectral analyses per pulsar, then reweight and refit the recovered posterior distribution to obtain the likelihood for the full array under arbitrary spectral models. Using the NUTS scheme and a GPU implementation, our method is more robust to the dimensionality of the posterior and we are able to efficiently model the GWB and intrinsic RN of every pulsar in the array with a free spectrum, without neglecting inter-frequency or -pulsar correlations, if desired.

If the stochastic background is realized via a population of a few loud SMBHBs, then it will exhibit higher-order statistical moments, beyond covariance~\citep{Higher_moments, Higher_moments2, Higher_moments3, GWB_dist1, GWB_dist3, GWB_dist2}. We will assume the Gaussian approximation, Eq.~(\ref{eq:a_prior}), is sufficient in the main body. Non-Gaussian features are discussed in Appendix~\ref{app:non-gauss}.

The hyper-parameters, $\boldsymbol{\eta}$, are subject to a hyper-prior. The power law amplitude will use a log-uniform hyper-prior, $\log_{10}A_\text{GWB}\sim\text{Uniform}(-20, -10)$, and the spectral index a uniform hyper-prior, $\gamma_\text{GWB}\sim\text{Uniform}(0, 7)$. Under a free spectral model, we'll use a log-uniform prior, $\log_{10}\rho_i\sim\text{Uniform}(-20, -5)$ for $i\in\{1,2,\dots,N_f\}$. Identical hyper-priors are used for the spectral models of intrinsic pulsar red noise \citep{RvH_PTAs_require_HBM}.

\subsection{Standardizing the pulsar timing posterior}\label{subsec:standardize_PTA}
The hierarchical PTA posterior, Eq.~(\ref{eq:PTA_HBM}), can be constructed by multiplying Eq.~(\ref{eq:likelihood}), Eq.~(\ref{eq:a_prior}), the prior on the timing model parameters, $\boldsymbol{\epsilon}_i\sim\mathcal{N}(0, \infty)$, and the hyper-prior on the parameters of the spectral model, $p(\boldsymbol{\eta})$. However, this is generally not the posterior sampled in standard analyses. Instead, the parameters of the linearized timing model are marginalized analytically from the hierarchical Bayesian model. This is useful not only because it reduces the dimension of the parameter space, but also because the timing model parameters are highly covariant with other signal and noise processes.

The analytic marginalization is equivalent to projecting the components of the analysis into a space orthogonal to the timing model, and is accomplished by replacing the inverse white noise covariance matrix $\mathbf{N}^{-1}\rightarrow\mathbf{\tilde{N}}^{-1}$, where
\begin{equation}\label{eq:Gmatrix_proj}
     \mathbf{\tilde{N}}^{-1} = \mathbf{G}(\mathbf{G}^\text{T}\mathbf{N}\mathbf{G})^{-1}\mathbf{G}^\text{T} 
\end{equation}
and $\mathbf{G}$ is built from the singular value decomposition (SVD) of the timing design matrix as in \cite{TMmarg2, TMmarg1, TMmarg3}. Alternatively, this projection can be accomplished using QR-decomposition,
\begin{equation}
    \mathbf{M} = \mathbf{Q}\mathbf{R} = [\mathbf{Q}_1,\,\mathbf{Q}_2]\begin{bmatrix}
        \mathbf{R}_1 \\
        \mathbf{0}
    \end{bmatrix}
\end{equation}
where $\mathbf{Q}$ is a $(n\times n)$ unitary matrix and $\mathbf{R}$ is a $(n\times m)$ upper triangular matrix with zeros populating the bottom $(n - m)$ rows, where $n$ and $m$ are the number of TOAs and timing model parameters in a particular pulsar, respectively. $\mathbf{Q}$ and $\mathbf{R}$ are partitioned such that $\mathbf{R}_1$ is a $(m\times m)$ upper triangular matrix, $\mathbf{0}$ is a $((n - m)\times m)$ zero matrix, and $\mathbf{Q}_1$ and $\mathbf{Q}_2$ with orthogonal columns are size $(n\times m)$ and $(n\times(n - m))$, respectively. Similar to the $\mathbf{G}$-matrix method, the columns of $\mathbf{Q}_2$ form an orthonormal basis for the subspace orthogonal to the timing model. Hence, $\mathbf{Q}_2^\text{T}\mathbf{M}=\mathbf{0}$ and we may replace $\mathbf{G}$ in Eq.~(\ref{eq:Gmatrix_proj}) with $\mathbf{Q}_2$. Using column pivoted QR-decomposition we may achieve this projection more efficiently than SVD and with greater numerical stability for the (nearly) rank-deficient timing model design matrix \cite{QRdecomp}.

The posterior, analytically marginalized over linear deviations to the timing model, is then
\begin{widetext}
\begin{equation}\label{eq:raw_posterior}
    p(\mathbf{a},\boldsymbol{\eta}|\boldsymbol{\delta t})\propto \frac{p(\boldsymbol{\eta})}{\sqrt{\text{det}(2\pi\tilde{\mathbf{N}})\text{det}(2\pi\boldsymbol{\phi})}}\,\text{exp}\bigg[-\frac{1}{2}\big(\boldsymbol{\delta t} - \mathbf{F}\mathbf{a}\big)^\text{T}\,\mathbf{\tilde{N}}^{-1}\,\big(\boldsymbol{\delta t} - \mathbf{F}\mathbf{a}\big)-\frac{1}{2}\mathbf{a}^\text{T}\,\boldsymbol{\phi}^{-1}\,\mathbf{a}\bigg]\,.
\end{equation}
\end{widetext}
While written in the notation of per-pulsar analyses as introduced in Sec.~\ref{sec:signal_comps}, it is understood that Eq.~(\ref{eq:raw_posterior}) models every pulsar in the array. That is, all constituent objects of the posterior above (and in what follows) are concatenated across pulsars in the array, except $\boldsymbol{\phi}$ whose definition, Eq.~(\ref{eq:prior_cov}), already spans every pulsar. e.g. $\boldsymbol{\delta t}\equiv[\boldsymbol{\delta t}_{(1)}, \boldsymbol{\delta t}_{(2)}, \dots, \boldsymbol{\delta t}_{(N_p)}]^\text{T}$ where $\boldsymbol{\delta t}_{(I)}$ denotes the TOAs of the $I^\text{th}$ pulsar, and $I\in\{1, 2,\dots, N_p\}$. With the white noise model fixed, evaluating Eq.~(\ref{eq:raw_posterior}) generally scales as $\mathcal{O}(N_p n^2)$ as it requires $N_p$ $(n\times n)$ matrix multiplications, assuming each of the $N_p$ pulsars has exactly $n$ TOAs.

Following the procedure of \cite{Lentati}, Eq.~(\ref{eq:raw_posterior}) can be written in a more efficient form
\begin{widetext}
\begin{equation}\label{eq:stochastic_posterior}
    p(\mathbf{a},\boldsymbol{\eta}|\boldsymbol{\delta t})\propto \frac{p(\boldsymbol{\eta})}{\sqrt{\text{det}(2\pi\boldsymbol{\phi})}}\,\text{exp}\bigg[-\frac{1}{2}\big(\mathbf{a} - \mathbf{\hat{a}}\big)^\text{T}\,\boldsymbol{\Sigma}^{-1}\,\big(\mathbf{a} - \mathbf{\hat{a}}\big) + \frac{1}{2}\mathbf{\hat{a}}^\text{T}\,\boldsymbol{\Sigma}^{-1}\,\mathbf{\hat{a}}\bigg]
\end{equation}
\end{widetext}
where $\mathbf{\hat{a}} = \boldsymbol{\Sigma}\mathbf{F}^\text{T}\mathbf{\tilde{N}}^{-1}\boldsymbol{\delta t}$, $\boldsymbol{\Sigma}^{-1}=\mathbf{F}^\text{T}\mathbf{\tilde{N}}^{-1}\mathbf{F} + \boldsymbol{\phi}^{-1}$, and we drop the $\boldsymbol{\delta t}^\text{T}\tilde{\mathbf{N}}^{-1}\boldsymbol{\delta t}$ term in the exponent and neglect the normalization factor $\text{det}(2\pi\tilde{\mathbf{N}})$. The white noise model is fixed in our analysis and these terms amount to constant multiplicative factors which do not influence the recovery of the target distribution. While mathematically equivalent, Eq.~(\ref{eq:stochastic_posterior}), is significantly more efficient than Eq.~(\ref{eq:raw_posterior}). $\mathbf{F}^\text{T}\,\mathbf{\tilde{N}}^{-1}\,\boldsymbol{\delta t}$ and $\mathbf{F}^\text{T}\,\mathbf{\tilde{N}}^{-1}\,\mathbf{F}$ can be computed once and stored for future evaluations, so Eq.~(\ref{eq:stochastic_posterior}) requires only matrix multiplications of size $(2N_fN_p\times 2N_fN_p)$. As the signal components of PTA datasets are red, relatively few low frequency bins are required to store signal information, and the Fourier domain is a compressed representation of PTA datasets, hence $N_f\ll n$. Posteriors that respect this compression, such as Eq.~(\ref{eq:stochastic_posterior}), will be more computationally efficient than posteriors using alternative representations.

At this stage, most analyses analytically integrate Eq.~(\ref{eq:stochastic_posterior}) with respect to the Fourier coefficients [c.f. \citep{Lentati}] marginalizing them from the model. The resulting density, while significantly lower-dimensional, contains a large $(N\times N)$ covariance matrix, where $N$ is the total number of TOAs for all pulsars in the array. This covariance matrix is dense due to inter-pulsar correlations from the GWB, and must be inverted for every posterior evaluation. The efficiency of the inversion is improved using the Woodbury matrix identity~\citep{Woodbury} but bottlenecks the analysis nonetheless. Rather than analytically marginalizing over the Fourier coefficients, we will keep Eq.~(\ref{eq:stochastic_posterior}) in its present form, and sample the Fourier coefficients numerically. Eq.~(\ref{eq:stochastic_posterior}) is computationally cheap to evaluate, requiring no expensive matrix inversions, but is high-dimensional due to the large number of Fourier coefficients. The inference on the spectral hyper-parameters $\boldsymbol{\eta}$, however, is equivalent to the analytic approach. It's simply a question of when the marginalization is performed: analytically at the posterior evaluation, or numerically via sampling and Monte Carlo integration.

The hierarchical posterior above, Eq.~(\ref{eq:stochastic_posterior}), is plagued by Neal's funnel. This can be seen in the conditional prior on the Fourier coefficients, Eq.~(\ref{eq:a_prior}). Say a power law spectral model is used to parameterize the covariance matrix, $\boldsymbol{\phi}$, a common modeling choice for PTA analyses. When the spectral index is large (small), the power is relatively constrained (free) in high frequency bins, and the respective coefficients to have small (large) variance. This forms a funnel analogous to that of Sec.~\ref{sec:toy_funnel}, but one which is significantly more difficult to sample directly, being high-dimensional across coefficients and hyper-parameters. The sharpness of the funnel may be reduced if a free spectral model is used to describe all red processes as in~\citep{escaping_funnel}. However, Eq.~(\ref{eq:stochastic_posterior}) is generally very difficult to sample directly which is why previous approaches have opted to analytically marginalize the Fourier coefficients from the model: not only is the dimension of the parameter space reduced, but the funnel is absent from the marginalized posterior geometry.

Rather than sampling Eq.~(\ref{eq:stochastic_posterior}) directly, we will sample the Fourier coefficients under a standardizing transform. This eases the difficulty of exploring Neal's funnel, while maintaining the hyper-efficient posterior evaluation. The high-dimensional parameter space is efficiently explored with HMC using a NUTS scheme. To perform a standardizing transform on the coefficients, we first estimate their mean and covariance. Examining Eq.~(\ref{eq:stochastic_posterior}), the mean and covariance of the Fourier coefficients are approximately $\mathbf{\hat{a}}$ and $\boldsymbol{\Sigma}$, respectively. This can be verified under the Laplace approximation by computing the \textit{maximum a posteriori} (MAP) solution for the Fourier coefficients, as in \cite{Lentati}, and identifying it with the mean of the distribution. The covariance can be estimated using the Hessian of the log-posterior: $-\partial_\mathbf{a}\partial_\mathbf{a}\ln p(\mathbf{a}, \boldsymbol{\eta}|\boldsymbol{\delta t})\vert_{\hat{\mathbf{a}}} = \boldsymbol{\Sigma}^{-1}$. The standardizing coordinate transform, Eq.~(\ref{eq:standardizing_transform}), for the Fourier coefficients in the PTA case is then
\begin{equation}\label{eq:PTA_standard_transform}
    (\mathbf{a}, \;\boldsymbol{\eta}) = T^{-1}(\mathbf{z}, \boldsymbol{\eta}) = (\mathbf{\hat{a}} + \mathbf{L}\mathbf{z},\;\boldsymbol{\eta})
\end{equation}
where $\mathbf{L}$ is the Cholesky decomposition of the covariance matrix, $\boldsymbol{\Sigma} = \mathbf{L}\mathbf{L}^\text{T}$. As in Sec.~\ref{sec:toy_funnel}, the mean and covariance used in the standardizing transform depend on the hyper-parameters, $\hat{\mathbf{a}}=\hat{\mathbf{a}}(\boldsymbol{\eta})$ and $\boldsymbol{\Sigma}=\boldsymbol{\Sigma}(\boldsymbol{\eta})$.

In practice, it is expensive to compute the Cholesky decomposition of the covariance matrix $\boldsymbol{\Sigma}$ which includes inter-pulsar correlations \footnote{Ironically, after using the Woodbury identity, the Cholesky decomposition of $\boldsymbol{\Sigma}$ is the computational bottleneck for evaluating the posterior in which the Fourier coefficients have been analytically marginalized. Our approach would be no faster than standard methods if we used the exact (inter-pulsar-correlated) Cholesky decomposition in the standardizing transform.}. However, the standardizing transformation need not be exact, and we may approximate the covariance in favor of a more computationally efficient standardizing transform. We choose to approximate the covariance matrix to that of a common uncorrelated red noise (CURN) process. That is, inter-pulsar correlations are neglected in the stochastic GWB model allowing us to factor the Cholesky decomposition per-pulsar. The Cholesky decomposition of the CURN covariance is $\boldsymbol{\Sigma}_\text{CURN} = (\mathbf{F}^\text{T}\mathbf{\tilde{N}}^{-1}\mathbf{F} + \boldsymbol{\phi}^{-1}_\text{CURN})^{-1} = \mathbf{L}_\text{CURN}\mathbf{L}_\text{CURN}^\text{T}$, where the prior covariance $\boldsymbol{\phi}_\text{CURN}$ is identical to that of Eq.~(\ref{eq:prior_cov}), but $\alpha_{IJ} = \delta_{IJ}$, neglecting inter-pulsar correlations. Similarly, we define $\hat{\mathbf{a}}_\text{CURN}=\boldsymbol{\Sigma}_\text{CURN}\mathbf{F}^\text{T}\tilde{\mathbf{N}}^{-1}\boldsymbol{\delta t}$.

While only an approximation, the CURN model is the primary component of the stochastic GWB, and effectively de-correlates the parameter space. The HD inter-pulsar correlation, Eq.~(\ref{eq:HD}), has a maximum correlation of 0.5 and is sub-dominant to the CURN model. In other words, Eq.~(\ref{eq:prior_cov}) is diagonal-dominant. This has been seen empirically in the latest published datasets such as the NANOGrav 15-year release, \cite{NG15}, which found a Bayes factor of $\sim10^{12}$ in favor of a model containing intrinsic pulsar RN and a CURN GWB over a model containing only pulsar noise. Meanwhile, a Bayes factor of $\sim10^2$ was found in favor of a HD correlated GWB over a CURN model, suggesting the GWB is dominated by diagonal CURN contributions, and only weakly influenced by off-diagonal inter-pulsar correlations. Hence the CURN model can be used in the standardizing transform to approximately de-correlate the parameter space.

The approximate and computationally efficient standardizing transform used in practice is then
\begin{equation}\label{eq:PTA_standard_transform_curn}
    (\mathbf{a}, \;\boldsymbol{\eta}) = T^{-1}(\mathbf{z}, \boldsymbol{\eta}) = (\mathbf{\hat{a}}_\text{CURN} + \mathbf{L}_\text{CURN}\,\mathbf{z},\;\boldsymbol{\eta})
\end{equation}
and the standardized density which is sampled instead of Eq.~(\ref{eq:stochastic_posterior}) is
\begin{widetext}
\begin{equation}\label{eq:standardized_stochastic_posterior}
    \tilde{p}(\mathbf{z},\boldsymbol{\eta}|\boldsymbol{\delta t})\propto \frac{p(\boldsymbol{\eta})\cdot\text{det}(\mathbf{L}_\text{CURN})}{\sqrt{\text{det}(2\pi\boldsymbol{\phi})}}\,\text{exp}\bigg[-\frac{1}{2}\big(\hat{\mathbf{a}}_\text{CURN} + \mathbf{L}_\text{CURN}\,\mathbf{z} - \mathbf{\hat{a}}\big)^\text{T}\,\boldsymbol{\Sigma}^{-1}\,\big(\hat{\mathbf{a}}_\text{CURN} + \mathbf{L}_\text{CURN}\,\mathbf{z} - \mathbf{\hat{a}}\big) + \frac{1}{2}\mathbf{\hat{a}}^\text{T}\,\boldsymbol{\Sigma}^{-1}\,\mathbf{\hat{a}}\bigg]\,,
\end{equation}
\end{widetext}
where we've been careful to include the determinant of the Jacobian of the transformation. In this case, the Jacobian of the transformation is the Cholesky decomposition of the CURN-approximated coefficient covariance matrix, $\boldsymbol{\Sigma}_\text{CURN} = \mathbf{L}_\text{CURN}\mathbf{L}_\text{CURN}^\text{T}$, whose determinant is extremely cheap to evaluate being a lower-triangular matrix (the determinant is simply the product of the diagonal elements). While we perform the coordinate transformation with respect to a CURN model, Eq.~(\ref{eq:standardized_stochastic_posterior}) enforces the inter-pulsar correlations through the presence of $\boldsymbol{\Sigma}$ so arbitrary inter-pulsar correlations (e.g. HD inter-pulsar correlations) can be modeled efficiently. Again the transformation, neglecting inter-pulsar correlations, does not perfectly de-correlate the parameter space. However, the estimated moments $\hat{\mathbf{a}}_\text{CURN}=\hat{\mathbf{a}}_\text{CURN}(\boldsymbol{\eta})$ and $\boldsymbol{\Sigma}_\text{CURN}=\boldsymbol{\Sigma}_\text{CURN}(\boldsymbol{\eta})$ are sufficient such that $\mathbf{z}$ approximately obeys a standard normal distribution. The true Fourier coefficients, $\mathbf{a}$, may be generated from our samples in $\mathbf{z}$ using Eq.~(\ref{eq:PTA_standard_transform_curn}).

Eq.~(\ref{eq:standardized_stochastic_posterior}) is the main result of this paper. It's worth noting that after sampling and mapping back to the original Fourier coefficients, our inference is identical to standard techniques. However, the standardized posterior does not require the inversion of any large dense matrices, retaining the hyper-efficient posterior formulation first presented in \citet{Lentati}. While standard posterior formulations analytically marginalize over the Fourier coefficients, we keep them as model parameters and sample them, performing the marginalization numerically. This means we sample over thousands of more parameters than standard approaches. As we've removed Neal's funnel via the coordinate transformation Eq.~(\ref{eq:PTA_standard_transform_curn}), the extra parameters $\mathbf{z}$ approximately obey a standard normal distribution. Such approximately normal high-dimensional distributions may be sampled extremely efficiently with HMC algorithms using fast automatic differentiation and XLA (Accelerated Linear Algebra) methods with a GPU-backend. The main results of this paper are obtained by implementing Eq.~(\ref{eq:standardized_stochastic_posterior}) in the \texttt{JAX}~\citep{JAX} package and performing HMC sampling with the \texttt{NumPyro}~\citep{NumPyro, NumPyro2} package on a GPU, see Section~\ref{sec:Results}.

\subsection{Generalizing to include deterministic signals}\label{subsec:gen2det}
It is straightforward to generalize the posterior, Eq.~(\ref{eq:raw_posterior}), to include deterministic signals. If the timing delays are modeled with a deterministic contribution $\mathbf{h}$, parameterized by $\boldsymbol{\theta}$, the posterior is modified
\begin{widetext}
\begin{equation}\label{eq:raw_det_posterior}
    p(\mathbf{a},\boldsymbol{\eta},\boldsymbol{\theta}|\boldsymbol{\delta t})\propto \frac{p(\boldsymbol{\eta})\cdot p(\boldsymbol{\theta})}{\sqrt{\text{det}(2\pi\mathbf{\tilde{N}})\text{det}(2\pi\boldsymbol{\phi})}}\,\text{exp}\bigg[-\frac{1}{2}\bigg(\boldsymbol{\delta t} - \mathbf{F}\mathbf{a} - \mathbf{h}(\boldsymbol{\theta})\bigg)^\text{T}\,\mathbf{\tilde{N}}^{-1}\,\bigg(\boldsymbol{\delta t} - \mathbf{F}\mathbf{a}-\mathbf{h}(\boldsymbol{\theta})\bigg)-\frac{1}{2}\mathbf{a}^\text{T}\,\boldsymbol{\phi}^{-1}\,\mathbf{a}\bigg]\,,
\end{equation}
\end{widetext}
where $p(\boldsymbol{\theta})$ is the prior on the parameters of the deterministic model. We will follow \cite{joint_CW} and express the deterministic model in a Fourier basis, so the deterministic delays are expressed as
\begin{equation}\label{eq:det_F_basis}
    \mathbf{h}(\boldsymbol{\theta}) = \mathbf{F}_\text{D}\,\mathbf{a}_\text{D}(\boldsymbol{\theta})\;\longleftrightarrow\;\mathbf{a}_\text{D}(\boldsymbol{\theta}) = \mathfrak{F}[\mathbf{h}(\boldsymbol{\theta})]\,,
\end{equation}
where $\mathbf{F}_\text{D}$ is the Fourier design matrix containing the basis and $\mathbf{a}_\text{D}$ the representation of the deterministic signal in a Fourier space. $\mathfrak{F}[\mathbf{h}(\boldsymbol{\theta})]$ denotes the discrete Fourier transform of the time-domain deterministic model, which may be performed analytically or numerically with a fast Fourier transform (FFT). A non-evolving continuous wave signal has a simple analytic Fourier transform, while more sophisticated deterministic delays may require numerical techniques.

The Fourier basis used for deterministic signals need not be identical to that of the stochastic components, $\mathbf{F}_\text{D}\neq\mathbf{F}$. In fact, it is advantageous to use a distinct basis as standard analyses define the Fourier basis for stochastic components with respect to the observation span of the array: the lowest frequency resolved is $1 / T$, where $T$ is the total duration of observations. As deterministic models are generally not periodic over this window, it is preferable to use an extended basis to avoid biases induced by Gibbs phenomena \citep{GibbsPhenomena} (see Appendix A of \citet{joint_CW} for a discussion).

Replacing the deterministic model with its Fourier representation, the generalized posterior, Eq.~(\ref{eq:raw_det_posterior}), can be written as
\begin{widetext}
    \begin{align}\label{eq:stoch_det_posterior}
        p(\mathbf{a},\boldsymbol{\eta},\boldsymbol{\theta}|\boldsymbol{\delta t})\propto \frac{p(\boldsymbol{\eta})\cdot p(\boldsymbol{\theta})}{\sqrt{\text{det}(2\pi\boldsymbol{\phi})}}\,\text{exp}\bigg[-&\frac{1}{2}\big(\mathbf{a}-\mathbf{\hat{a}})^\text{T}\,\boldsymbol{\Sigma}^{-1}\,(\mathbf{a}-\mathbf{\hat{a}}) + \frac{1}{2}\mathbf{\hat{a}}^\text{T}\,\boldsymbol{\Sigma}^{-1}\,\mathbf{\hat{a}} \nonumber \\
        &+ \boldsymbol{\delta t}^\text{T}\,\mathbf{\tilde{N}}^{-1}\,\mathbf{F}_\text{D}\,\mathbf{a}_\text{D} - \mathbf{a}^\text{T}\,\mathbf{F}^\text{T}\,\mathbf{\tilde{N}}^{-1}\,\mathbf{F}_\text{D}\,\mathbf{a}_\text{D}-\frac{1}{2}\mathbf{a}_\text{D}^\text{T}\,\mathbf{F}_\text{D}^\text{T}\,\mathbf{\tilde{N}}^{-1}\,\mathbf{F}_\text{D}\,\mathbf{a}_\text{D}\bigg]\,.
    \end{align}
\end{widetext}
Differing from those of the stochastic model, the Fourier coefficients for the deterministic model are not sampled directly or under a bijective coordinate transformation. Instead the parameters of the deterministic model, $\boldsymbol{\theta}$, are sampled and the coefficients are the Fourier transform of the corresponding time-domain deterministic signal, $\mathbf{h}=\mathbf{h}(\boldsymbol{\theta})$, so there exists the mapping $\mathbf{a}_\text{D} = \mathbf{a}_\text{D}(\boldsymbol{\theta})$. In similar fashion to the pure stochastic model above, Eq.~(\ref{eq:stoch_det_posterior}) is more computationally efficient to evaluate than the equivalent Eq.~(\ref{eq:raw_det_posterior}). Again, this is because large matrix multiplications over all TOAs are replaced with multiplications over the compressed Fourier basis, and relevant inner products (e.g. $\boldsymbol{\delta t}^\text{T}\,\mathbf{\tilde{N}}^{-1}\,\mathbf{F}_\text{D}$, $\mathbf{F}^\text{T}\,\mathbf{\tilde{N}}^{-1}\,\mathbf{F}_\text{D}$, etc.) are computed once and stored for future evaluations.

Another computational overhead is the conversion of the deterministic model into a Fourier representation. The numerical FFT is generally computationally cheaper than Eq.~(\ref{eq:raw_det_posterior}), where the deterministic model must be evaluated over every observed TOA. Say the same frequency resolution, $N_f$, is desired in the deterministic model as in the stochastic model. Then we need only query the deterministic model $\mathcal{O}(N_f)$ times and a FFT yields the frequency representation in $\mathcal{O}(N_f\log N_f)$ operations. Generally $N_f\ll n$ in PTA analysis, so obtaining the frequency-domain representation of the deterministic signal is not a significant computational cost, relative to the cost of matrix multiplications in the posterior.

In the case $\mathbf{F}_\text{D}=\mathbf{F}$, Eq.~(\ref{eq:stoch_det_posterior}) reduces to
\begin{widetext}
    \begin{equation}\label{eq:FD_F_stoch_det_posterior}
        p(\mathbf{a},\boldsymbol{\eta},\boldsymbol{\theta}|\boldsymbol{\delta t})\propto \frac{p(\boldsymbol{\eta})\cdot p(\boldsymbol{\theta})}{\sqrt{\text{det}(2\pi\boldsymbol{\phi})}}\,\text{exp}\bigg[-\frac{1}{2}\big(\mathbf{a}+\mathbf{a}_\text{D}-\mathbf{\hat{a}})^\text{T}\,\boldsymbol{\Sigma}^{-1}\,(\mathbf{a}+\mathbf{a}_\text{D}-\mathbf{\hat{a}}) + \frac{1}{2}\mathbf{\hat{a}}^\text{T}\,\boldsymbol{\Sigma}^{-1}\,\mathbf{\hat{a}} + \mathbf{a}^\text{T}\,\boldsymbol{\phi}^{-1}\,\mathbf{a}_\text{D} + \frac{1}{2}\mathbf{a}^\text{T}_\text{D}\,\boldsymbol{\phi}^{-1}\,\mathbf{a}_\text{D}\bigg]\,,
    \end{equation}
\end{widetext}
which is similar to the posterior for a purely stochastic model, Eq.~(\ref{eq:stochastic_posterior}). The extra additive terms in the exponent are corrections so the deterministic signal is not influenced hierarchically by the spectral model, which is solely intended for stochastic contributions.

We may generalize the standardizing transform presented for the stochastic model above, Eq.~(\ref{eq:PTA_standard_transform}), to include contributions from deterministic signals. The goal is to estimate the (conditional) mean and covariance of the stochastic Fourier coefficients from the posterior generalized to include deterministic components, Eq.~(\ref{eq:stoch_det_posterior}). The presence of a deterministic signal shifts the MAP solution from $\mathbf{\hat{a}}$, of the purely stochastic case, to
\begin{equation}\label{eq:det_mean}
    \mathbf{\bar{a}} = \mathbf{\hat{a}} - \boldsymbol{\Sigma}\mathbf{F}^\text{T}\,\mathbf{\tilde{N}}^{-1}\,\mathbf{F}_\text{D}\,\mathbf{a}_\text{D} = \boldsymbol{\Sigma}\mathbf{F}^\text{T}\,\mathbf{\tilde{N}}^{-1}(\boldsymbol{\delta t} - \mathbf{h})
\end{equation}
which can be derived from the MAP condition, $\partial_\mathbf{a}\ln p(\mathbf{a},\boldsymbol{\eta},\boldsymbol{\theta})\big|_{\mathbf{a}=\mathbf{\bar{a}}}=\mathbf{0}$. Intuitively, to include deterministic contributions, one needs only subtract the deterministic signal from the data, then compute the MAP solution as in Sec.~\ref{subsec:standardize_PTA}. The covariance of the stochastic Fourier coefficients is unchanged by the presence of deterministic signals as seen by the Hessian $-\partial_\mathbf{a}\partial_\mathbf{a}\ln p(\mathbf{a},\boldsymbol{\eta},\boldsymbol{\theta}) = \boldsymbol{\Sigma}^{-1}$.

As in Sec.~\ref{subsec:standardize_PTA}, we will approximate the standardizing transform by neglecting inter-pulsar correlations with the CURN model. The generalization of Eq.~(\ref{eq:PTA_standard_transform_curn}) to include deterministic contributions is
\begin{equation}\label{eq:decenter_det_PTA}
    (\mathbf{a}, \;\boldsymbol{\eta},\;\boldsymbol{\theta}) = T^{-1}(\mathbf{z}, \boldsymbol{\eta},\boldsymbol{\theta}) = (\mathbf{\bar{a}}_\text{CURN} + \mathbf{L}_\text{CURN}\,\mathbf{z},\;\boldsymbol{\eta},\;\boldsymbol{\theta})
\end{equation}
where $\bar{\mathbf{a}}_\text{CURN}=\mathbf{\hat{a}}_\text{CURN} - \boldsymbol{\Sigma}_\text{CURN}\mathbf{F}^\text{T}\,\mathbf{\tilde{N}}^{-1}\,\mathbf{F}_\text{D}\,\mathbf{a}_\text{D}$. While mapped trivially by the coordinate transformation, the hyper-parameters and parameters of the deterministic model determine the standardizing transformation performed on the coefficients, $\bar{\mathbf{a}}_\text{CURN}=\bar{\mathbf{a}}_\text{CURN}(\boldsymbol{\eta},\boldsymbol{\theta})$ and $\boldsymbol{\Sigma}_\text{CURN}=\boldsymbol{\Sigma}_\text{CURN}(\boldsymbol{\eta})$. The standardized posterior sampled is
\begin{widetext}
    \begin{align}\label{eq:standardized_det_posterior}
        \tilde{p}(\mathbf{z},\boldsymbol{\eta},\boldsymbol{\theta}|\boldsymbol{\delta t})\propto \frac{p(\boldsymbol{\eta})\cdot p(\boldsymbol{\theta})\cdot\text{det}(\mathbf{L}_\text{CURN})}{\sqrt{\text{det}(2\pi\boldsymbol{\phi})}}\,\text{exp}\bigg[&-\frac{1}{2}\big(\mathbf{\bar{a}}_\text{CURN} + \mathbf{L}_\text{CURN}\,\mathbf{z}-\mathbf{\hat{a}})^\text{T}\,\boldsymbol{\Sigma}^{-1}\,(\mathbf{\bar{a}}_\text{CURN} + \mathbf{L}_\text{CURN}\,\mathbf{z}-\mathbf{\hat{a}}) \nonumber \\
        &+ \frac{1}{2}\mathbf{\hat{a}}^\text{T}\,\boldsymbol{\Sigma}^{-1}\,\mathbf{\hat{a}} + \boldsymbol{\delta t}^\text{T}\,\mathbf{\tilde{N}}^{-1}\,\mathbf{F}_\text{D}\,\mathbf{a}_\text{D} -\frac{1}{2}\mathbf{a}_\text{D}^\text{T}\,\mathbf{F}_\text{D}^\text{T}\,\mathbf{\tilde{N}}^{-1}\,\mathbf{F}_\text{D}\,\mathbf{a}_\text{D}\nonumber \\
        &- (\mathbf{\bar{a}}_\text{CURN} + \mathbf{L}_\text{CURN}\,\mathbf{z})^\text{T}\,\mathbf{F}^\text{T}\,\mathbf{\tilde{N}}^{-1}\,\mathbf{F}_\text{D}\,\mathbf{a}_\text{D}\bigg]\,.
    \end{align}
\end{widetext}
Again, we're careful to include the determinant of the Jacobian of the transformation. As in Sec.~\ref{subsec:standardize_PTA}, the standardizing transformation does not perfectly de-correlate the parameter space, but is computationally efficient and a sufficient estimate so the coefficients $\mathbf{z}$ approximately obey a standard normal distribution. While the standardizing transformation neglects inter-pulsar correlations, Eq.~(\ref{eq:standardized_det_posterior}) does include such correlations so our inference is identical to standard analyses. The original Fourier coefficients can be generated from the transformed samples using Eq.~(\ref{eq:decenter_det_PTA}).

\subsection{Generalizing to include inter-frequency correlations}\label{subsec:inter-freq}
Most spectral models, including the power law Eq.~(\ref{eq:power_law}) and free spectral model, assume the Fourier modes are uncorrelated. However, because PTA data is unevenly sampled in the time-domain, the inferred Fourier coefficients will inevitably be correlated \citep{Lentati, vH_Gaussian_advances}. Moreover, as the signal and noise processes we measure persist longer than our finite observation span, a rectangular window has effectively been applied to the data. The Fourier representation of the signal is then the convolution of the original Fourier expansion with the Fourier transform of the window function - a cardinal sine function, inducing additional correlations between the Fourier modes \citep{BeyondDiag, OptHDcorr, Legendre_basis}. The Fourier modes were assumed orthogonal in previous sections, Eq.~(\ref{eq:prior_cov}) being diagonal in frequency-space. This section illustrates how our coefficient sampling method is compatible with non-diagonal prior covariance matrices.

In the evaluation of the posterior, Eq.~(\ref{eq:stochastic_posterior}), we must compute the inverse of the $(2N_fN_p\times2N_fN_p)$ prior covariance matrix, $\boldsymbol{\phi}$. When inter-frequency correlations are neglected, this can be accomplished efficiently by batching our inverse over frequency bins, and inverting $2N_f$ $(N_p\times N_p)$ matrices in parallel. This is not possible when including inter-frequency correlations because $\boldsymbol{\phi}$ is dense in inter-pulsar \textit{and} -frequency correlations, and a straightforward inversion is computationally expensive. To avoid this bottleneck, we will separate the GWB and pulsar RN into separate Gaussian processes as in Appendix~\ref{app:split_coeffs}, where the prior covariance matrix for the coefficients which represent the background and intrinsic pulsar noise is $\boldsymbol{\phi}_\text{GWB}$ and $\boldsymbol{\phi}_\text{RN}$, respectively. The posterior for this parameterization is derived in Appendix~\ref{app:split_coeffs} at Eq.~(\ref{eq:separated_post}). The evaluation of this posterior requires us to invert both $\boldsymbol{\phi}_\text{RN}$ and $\boldsymbol{\phi}_\text{GWB}$, but we can compute these inverses more efficiently than that of the combined covariance matrix.

As the intrinsic pulsar noise, by definition, is independent across pulsars, $\boldsymbol{\phi}_\text{RN}$ is block-diagonal (or rather diagonal in pulsar-space, but dense in frequency-space due to windowing effects). We can therefore compute $\boldsymbol{\phi}_\text{RN}^{-1}$ by batching our matrix inversion across pulsars, and inverting $N_p$ $(2N_f\times2N_f)$ matrices in parallel. To compute $\boldsymbol{\phi}_\text{GWB}^{-1}$ efficiently, we note the GWB contribution to Eq.~(\ref{eq:prior_cov}) can be written as $\boldsymbol{\phi}_\text{GWB} = \boldsymbol{\Gamma}\otimes\boldsymbol{\varphi}$, where $\boldsymbol{\Gamma}$ is the overlap reduction function encoding inter-pulsar correlations and $\boldsymbol{\varphi}$ the common power spectrum across pulsars. That is, $\boldsymbol{\phi}_\text{GWB}$ remains dense in pulsar-space (due to $\Gamma$ being $(N_p\times N_p)$ dense) and frequency-space (due to $\boldsymbol{\varphi}$ being $(2N_f\times2N_f)$ dense). Thanks to its Kronecker structure, the inverse may still be computed cheaply, $\boldsymbol{\phi}_\text{GWB}^{-1}=\boldsymbol{\Gamma}^{-1}\otimes\boldsymbol{\varphi}^{-1}$, requiring only the inversion of one $(N_p\times N_p)$ and one $(2N_f\times2N_f)$ matrix.

We are now in a position to sample Eq.~(\ref{eq:separated_post}), with $\boldsymbol{\phi}_\text{GWB/RN}$ including inter-frequency correlations, under the standardizing transformation Eq.~(\ref{eq:split_standardizing_transform}). However, recall in computing the inverse prior covariance matrix $\boldsymbol{\phi}_\text{RN}$ we must invert $N_p$ dense $(2N_f\times2N_f)$ matrices. This is precisely the same computational cost as analytically marginalizing over the Fourier coefficients which represent the intrinsic pulsar red noise. Therefore, we might as well perform this analytic marginalization so we only have to numerically sample the remaining coefficients representing the GWB. This choice also reduces the computational cost of the standardizing transform. Completing the square, Eq.~(\ref{eq:separated_post}) is rewritten
\begin{widetext}
\begin{align}
    p(\mathbf{a}_\text{GWB},\mathbf{a}_\text{RN},\boldsymbol{\eta}|\boldsymbol{\delta t})\propto&\frac{p(\boldsymbol{\eta})}{\sqrt{\text{det}(2\pi\tilde{\mathbf{N}})\cdot\text{det}(2\pi\boldsymbol{\phi}_\text{GWB})\cdot\text{det}(2\pi\boldsymbol{\phi}_\text{RN})}} \\ \nonumber
    &\times \text{exp}\bigg[-\frac{1}{2}(\mathbf{a}_\text{RN}-\grave{\mathbf{a}}_\text{RN})^\text{T}\,\boldsymbol{\Sigma}_\text{RN}^{-1}\,(\mathbf{a}_\text{RN}-\grave{\mathbf{a}}_\text{RN}) + \frac{1}{2}\grave{\mathbf{a}}_\text{RN}^\text{T}\,\boldsymbol{\Sigma}_\text{RN}^{-1}\,\grave{\mathbf{a}}_\text{RN}\bigg] \\ \nonumber
    &\times \text{exp}\bigg[-\frac{1}{2}\mathbf{a}_\text{GWB}^\text{T}\,\boldsymbol{\Sigma}_\text{GWB}^{-1}\,\mathbf{a}_\text{GWB} + \boldsymbol{\delta t}^\text{T}\,\tilde{\mathbf{N}}^{-1}\,\mathbf{F}\,\mathbf{a}_\text{GWB}\bigg]
\end{align}
\end{widetext}
where as in Appendix~\ref{app:split_coeffs} $\boldsymbol{\Sigma}_\text{GWB/RN} = (\mathbf{F}^\text{T}\tilde{\mathbf{N}}^{-1}\mathbf{F} + \boldsymbol{\phi}_\text{GWB/RN}^{-1})^{-1}$ and $\grave{\mathbf{a}} = \boldsymbol{\Sigma}_\text{RN}(\mathbf{F}^\text{T}\tilde{\mathbf{N}}^{-1}\boldsymbol{\delta t} - \mathbf{F}^\text{T}\tilde{\mathbf{N}}^{-1}\mathbf{F}\mathbf{a}_\text{GWB})$ represents the MAP coefficients representing intrinsic pulsar noise, conditioned on the coefficients representing the background. In this form, the posterior is Gaussian in the red noise coefficients and we may analytically marginalize them from the model. The resulting posterior is
\begin{widetext}
\begin{align}\label{eq:post_interfreq_marg}
    p(\mathbf{a}_\text{GWB},\boldsymbol{\eta}|\boldsymbol{\delta t})&\propto \int p(\mathbf{a}_\text{GWB},\mathbf{a}_\text{RN},\boldsymbol{\eta}|\boldsymbol{\delta t})\,d\mathbf{a}_\text{RN} \\ \nonumber
    &\propto \sqrt{\frac{\text{det}(\boldsymbol{\Sigma}_\text{RN})}{\text{det}(\tilde{\mathbf{N}})\cdot\text{det}(\boldsymbol{\phi}_\text{RN})\cdot\text{det}(\boldsymbol{\phi}_\text{GWB})}}\times\text{exp}\bigg[-\frac{1}{2}\mathbf{a}_\text{GWB}^\text{T}\,\boldsymbol{\Sigma}^{-1}_\text{GWB}\,\mathbf{a}_\text{GWB} + \boldsymbol{\delta t}^\text{T}\,\tilde{\mathbf{N}}^{-1}\,\mathbf{F}\,\mathbf{a}_\text{GWB} \\ \nonumber
    &\hspace{80mm}+\frac{1}{2}\grave{\mathbf{a}}_\text{RN}^\text{T}\,\boldsymbol{\Sigma}_\text{RN}\,\grave{\mathbf{a}}_\text{RN}\bigg]\,.
\end{align}
\end{widetext}

To sample the remaining Fourier coefficients which represent the GWB efficiently, we derive the appropriate standardizing transformation. The mean and covariance (estimated under the Laplace approximation as above) are $\grave{\mathbf{a}}_\text{GWB} = \boldsymbol{\Phi}_\text{GWB}(\mathbf{F}^\text{T}\tilde{\mathbf{N}}^{-1}\boldsymbol{\delta t} - \mathbf{F}^\text{T}\tilde{\mathbf{N}}^{-1}\mathbf{F}\boldsymbol{\Sigma}_\text{RN}\mathbf{F}^\text{T}\tilde{\mathbf{N}}^{-1}\boldsymbol{\delta t})$ and $\boldsymbol{\Phi}_\text{GWB} = (\mathbf{F}^\text{T}\tilde{\mathbf{N}}^{-1}\mathbf{F}+\boldsymbol{\phi}^{-1}_\text{GWB} - \mathbf{F}^\text{T}\tilde{\mathbf{N}}^{-1}\mathbf{F}\boldsymbol{\Sigma}_\text{RN}\mathbf{F}^\text{T}\tilde{\mathbf{N}}^{-1}\mathbf{F})^{-1}$, respectively. The standardizing transformation, Eq.~(\ref{eq:PTA_standard_transform}) is modified under the replacement $\hat{\mathbf{a}}\rightarrow\grave{\mathbf{a}}$ and $\boldsymbol{\Sigma}\rightarrow\boldsymbol{\Phi}_\text{GWB}$ to
\begin{equation}\label{eq:standard_transform_interfreq}
    (\mathbf{a}_\text{GWB},\boldsymbol{\eta}) = T^{-1}(\mathbf{z}, \boldsymbol{\eta}) = (\grave{\mathbf{a}}_\text{GWB} + \mathbf{L}\mathbf{z}, \boldsymbol{\eta)}
\end{equation}
where $\mathbf{L}$ is the Cholesky decomposition of the covariance matrix $\boldsymbol{\Phi}_\text{GWB}$ under the CURN approximation. We do, however, include inter-frequency correlations in the construction of the transformation.

\begin{figure*}
    \centering
    \includegraphics[width=0.97\textwidth]{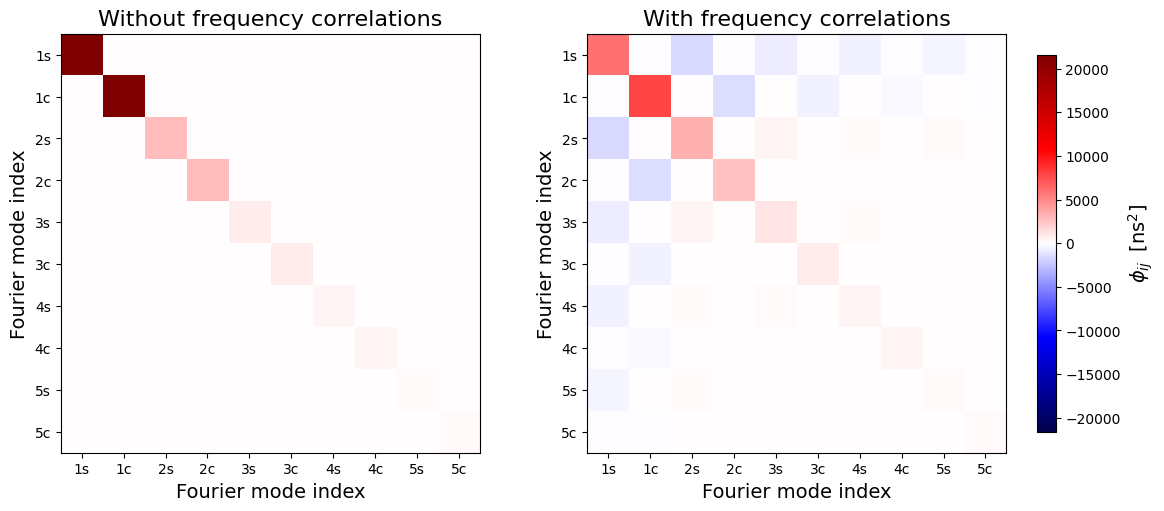}
    \caption{Frequency-domain prior covariance matrices, $\boldsymbol{\phi}$, with and without frequency correlations from window effects. Both are evaluated at the same power law parameters $\log_{10}A = -14.5$ and $\gamma = 3.0$. The tick labels on the x- and y-axes correspond to the frequency bin and corresponding Fourier mode, e.g. ``3c" corresponds to the cosine mode amplitude in the third frequency bin and ``3s" the sine amplitude of the same bin.}
    \label{fig:cov}
\end{figure*}

In practice, the inter-frequency correlated prior covariance matrix can be numerically unstable or computationally costly to compute. However, the \texttt{FFTInt} approach of \cite{BeyondDiag} approximates the time-domain covariance matrix efficiently and accurately by interpolating the matrix from a sparse grid of regular time samples to the observed TOAs. We adopt this approach and obtain the necessary frequency-domain covariance matrix, $\boldsymbol{\phi}$, using a two-dimensional FFT. The frequency-domain covariance with and without frequency correlations from window effects evaluated using a power law spectral model is shown in Fig.~\ref{fig:cov}. Estimating inter-frequency correlations requires the spectral model to have a continuous power spectral density function (e.g. a power law is continuous Eq.~(\ref{eq:power_law})). Other models like the free spectral model are not endowed with a continuous definition. Nonetheless, we may approximate the inter-frequency correlations under these models using a spline or some interpolation scheme to achieve a continuum limit.

In summary, Eq.~(\ref{eq:post_interfreq_marg}) is a posterior, equivalent to Eq.~(\ref{eq:stochastic_posterior}) after marginalizing over the Fourier coefficients representing intrinsic pulsar noise, but one which can include inter-frequency correlations efficiently. Eq.~(\ref{eq:stochastic_posterior}) cannot efficiently include inter-frequency correlations because computing $\boldsymbol{\phi}^{-1}$ is costly as it is a large dense matrix, with inter-pulsar and -frequency correlations. By separating the Gaussian processes which represent the intrinsic pulsar RN and GWB, we may compute $\boldsymbol{\phi}_\text{RN/GWB}^{-1}$ efficiently in the evaluation of Eq.~(\ref{eq:post_interfreq_marg}). The frequency-domain prior covariance matrix is computed with the two-dimensional FFT of the corresponding time-domain matrix from the \texttt{FFTInt} method. Our derivation above did not include deterministic signals as in Sec.~\ref{subsec:gen2det}. Nonetheless, our results Eq.~(\ref{eq:post_interfreq_marg}) and Eq.~(\ref{eq:standard_transform_interfreq}) can be generalized the include deterministic signals under the simple replacement $\boldsymbol{\delta t} \rightarrow \boldsymbol{\delta t} - \mathbf{h} = \boldsymbol{\delta t} - \mathbf{F}_\text{D}\mathbf{a}_D(\theta)$.

\section{Implementation techniques}
\subsection{Single precision}
The efficiency of our methods is significantly improved by operating entirely in single precision. That is, floating-point numbers are represented with 32 bits, as opposed to 64 bits (double precision) which is the default for many \texttt{Python} environments. Roughly speaking, this decision reduces the number of reliable digits from 15 to 7. We use units of nano-seconds (ns) rather than seconds (s) to faithfully represent the data. In nano-seconds, timing residuals are $\mathcal{O}(10^2)$ and more numerically stable than the alternative in seconds which is $\mathcal{O}(10^{-7})$. This convention is used for all objects in the posterior. For example, covariance matrices use units of ($\text{ns}^2$). Moreover, we drop the weighting of the determinant of the projected white noise covariance matrix, $\text{det}(2\pi\tilde{\mathbf{N}})$, in the posterior evaluation from Eq.~(\ref{eq:raw_posterior}) to Eq.~(\ref{eq:stochastic_posterior}) so evaluations of the posterior may be resolved in single precision. As we fix the white noise model during the analysis, this amounts to neglecting a constant scaling in our posterior and does not affect our inference.

In gamma-ray PTAs \citep{gamma-ray-PTA}, the matrix $\mathbf{F}^\text{T}\tilde{\mathbf{N}}^{-1}\mathbf{F}$ is singular, and the likelihood not normalizable. To remedy this, we may regularize the posterior, Eq.~(\ref{eq:stochastic_posterior}), as in \citet{Regularizing_PTA}, so that up to an overall scaling
\begin{widetext}
    \begin{equation}\label{eq:regularize}
    p(\mathbf{a},\boldsymbol{\eta}|\boldsymbol{\delta t})\propto p(\boldsymbol{\eta})\,\sqrt{\frac{\text{det}(2\pi\boldsymbol{\phi}_0)}{\text{det}(2\pi\boldsymbol{\phi})}}\,\text{exp}\bigg[-\frac{1}{2}\big(\mathbf{a} - \mathbf{\hat{a}}_0\big)^\text{T}\,\boldsymbol{\Sigma}^{-1}_0\,\big(\mathbf{a} - \mathbf{\hat{a}}_0\big) - \frac{1}{2}\mathbf{a}^\text{T}\,\big(\boldsymbol{\phi}^{-1} - \boldsymbol{\phi}^{-1}_0\big)\,\mathbf{a} + \frac{1}{2}\hat{\mathbf{a}}_0\,\boldsymbol{\Sigma}^{-1}_0\,\hat{\mathbf{a}}_0\bigg]
    \end{equation}
\end{widetext}
where $\hat{\mathbf{a}}_0 = \boldsymbol{\Sigma}_0\mathbf{F}^\text{T}\tilde{\mathbf{N}}^{-1}\boldsymbol{\delta t}$. $\boldsymbol{\Sigma}_0$ and $\boldsymbol{\phi}_0$ are \textit{reference} covariance matrices, identical to $\boldsymbol{\Sigma}$ and $\boldsymbol{\phi}$, respectively, but evaluated at a set of reference hyper-parameters, $\boldsymbol{\eta}_0$, and held fixed throughout the analysis. Note that Eq.~(\ref{eq:regularize}) is identical to Eq.~(\ref{eq:stochastic_posterior}), up to an overall scaling in the case of radio PTAs. However in gamma-ray PTAs, Eq.~(\ref{eq:stochastic_posterior}) is ill-conditioned whereas Eq.~(\ref{eq:regularize}) is a normalizable probability density.

We must also take care to implement deterministic models in single precision. If the CW model from \cite{CWsPTA, CWsPTA2, EllisCW, EllisCWs2} is implemented naively in single precision, the model is prone to catastrophic cancellation. This is seen in the phase-evolution of the CW signal,
\begin{equation}\label{eq:phase_evolve}
    \Phi(t)  = \Phi_0 + \frac{1}{32 {\mathcal M}^{5/3}}  \left( \omega_0^{-\frac{5}{3}} - \omega^{-\frac{5}{3}} \right)\,
\end{equation}
where $\Phi_0$ and $\omega_0$ are the reference phase and frequency of the CW, respectively. $\mathcal{M}$ is the chirp mass and $\omega$ the frequency of the CW. The frequency itself evolves as
\begin{equation}\label{eq:CW_evolve}
    \omega(t) = \omega_0\bigg(1 - \frac{256}{5}\mathcal{M}^{5/3}\omega_0^{8/3}t\bigg)^{-3/8}\,,
\end{equation}
where $t$ the time of evolution. In some regions of parameter space, the difference between the frequency and its reference value in Eq.~(\ref{eq:phase_evolve}) is not resolvable in single precision. In such cases, we substitute Eq.~(\ref{eq:CW_evolve}) into Eq.~(\ref{eq:phase_evolve}) and express the phase of evolution of the CW as
\begin{align}\label{eq:phase_evolve_Taylor}
    \Phi(t) &= \Phi_0 + \frac{1}{32(\mathcal{M}\omega_0)^{5/3}}\bigg[1 - (1 - x)^{5/8}\bigg] \\
    &\approx \Phi_0 + \frac{x}{(\mathcal{M}\omega_0)^{5/3}}\,,
\end{align}
where $x\equiv\mathcal{M}^{5/3}\omega_0^{8/3}t\ll 1$. Similar situations arise in the amplitude and pulsar term calculations of the CW model where we apply an identical treatment, only keeping resolvable terms in the Taylor series.

\subsection{Batching}\label{subsec:batching}
While Eq.~(\ref{eq:standardized_stochastic_posterior}) and the analogous regularized Eq.~(\ref{eq:regularize}) are convenient forms for the posterior in which the mean, covariance, and standardizing transform can be ``read-off", it is more computationally efficient to implement them in the numerically equivalent form
\begin{widetext}
    \begin{equation}\label{eq:posterior_implemented}
        \tilde{p}(\mathbf{z}, \boldsymbol{\eta}|\boldsymbol{\delta t}) \propto \frac{p(\boldsymbol{\eta})\cdot\text{det}(\mathbf{L}_\text{CURN})}{\sqrt{\text{det}(2\pi\boldsymbol{\phi})}}\,\text{exp}\bigg[\boldsymbol{\delta t}^\text{T}\tilde{\mathbf{N}}^{-1}\mathbf{F}\mathbf{a} - \frac{1}{2}\mathbf{a}^\text{T}\mathbf{F}^\text{T}\tilde{\mathbf{N}}^{-1}\mathbf{F}\mathbf{a}-\frac{1}{2}\mathbf{a}^\text{T}\boldsymbol{\phi}^{-1}\mathbf{a}\bigg]
    \end{equation}
\end{widetext}
where via the standardizing transform $\mathbf{a} = \hat{\mathbf{a}} + \mathbf{L}_\text{CURN}\mathbf{z}$. In this form, all terms without $\boldsymbol{\phi}$ are per-pulsar (including the standardizing transform which uses $\boldsymbol{\phi}_\text{CURN}$), and we may batch (or parallelize) our computations over pulsars using a GPU for extremely efficient evaluation. $\boldsymbol{\phi}$ includes inter-pulsar correlations, but does not include frequency correlations, and we may calculate its inverse, determinant, and derived matrix products by batching over frequencies (i.e. by inverting $2N_f$ $(N_p\times N_p)$ matrices in parallel to obtain its inverse) for efficient evaluation on a GPU. A posterior which includes inter-pulsar and -frequency correlations and the associated standardizing coordinate transformation was presented in Sec.~\ref{subsec:inter-freq}.

\begin{figure*}
    \centering
    \includegraphics[width=0.98\linewidth]{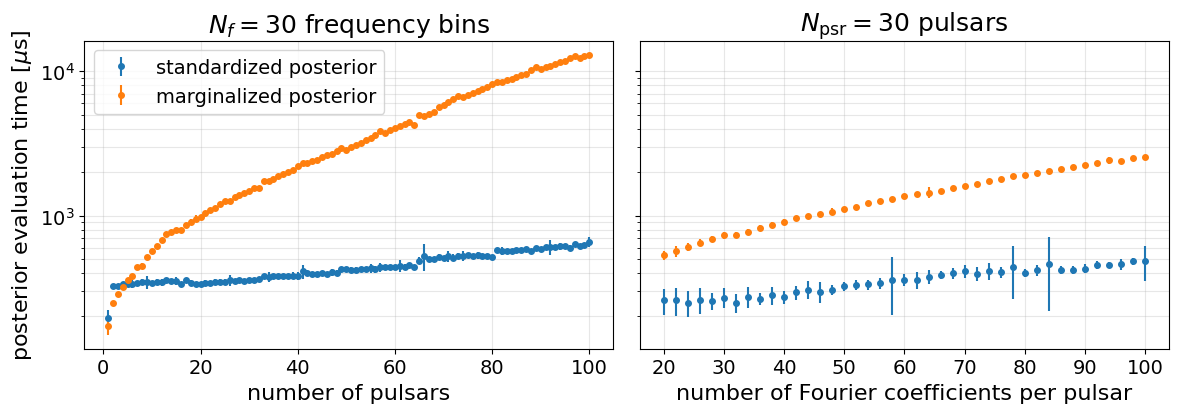}
    \caption{Evaluation times of the posterior probability density function on an NVIDIA GeForce RTX 3090 GPU, as function of the number of pulsars and frequency bins in the array. The results in blue denote the evaluation times for the standardized posterior, Eq.~(\ref{eq:posterior_implemented}), and the results in orange are the evaluation times for the posterior which has the Fourier coefficients analytically marginalized, as in \cite{Lentati}. Each pulsar used to construct the timing dataset was simulated and observed approximately monthly for 15 years.}
    \label{fig:timing}
\end{figure*}

Thanks to GPU parallelization, evaluating Eq.~(\ref{eq:posterior_implemented}) and its analogous forms which include regularization and/or deterministic signal contributions scales sub-linearly with the number of pulsars. Alternatively, if the Fourier coefficients had been analytically marginalized from the model the posterior evaluation scales worse than quadratically with the number of pulsars. We implement and time both the standardized posterior Eq.~(\ref{eq:posterior_implemented}) and the posterior marginalized over the Fourier coefficients (as derived in \cite{Lentati}) on an NVIDIA GeForce RTX 3090 GPU as a function of the number of pulsars and frequency bins. Each pulsar used in the timing was simulated and observed approximately monthly for 15 years. The results are shown in Fig.~\ref{fig:timing}. We estimate the polynomial scaling law of the posterior evaluation time with the number of pulsars and frequency bins, for $N_p>60$ and $N_f>20$. The standardized posterior scales as $\sim\mathcal{O}(N_p^{0.7})$ and $\sim\mathcal{O}(N_f^{0.6})$ with respect to the number of pulsars and frequency bins, respectively. The analytically marginalized posterior goes as $\sim\mathcal{O}(N_p^{2.4})$ and $\sim\mathcal{O}(N_f^{1.2})$ with respect to the number of pulsars and frequency bins, respectively. While the standardized posterior requires us to sample thousands of additional parameters, they approximately obey uncorrelated standard normal distributions and are sampled efficiently with HMC.

\section{Analyses of real and simulated datasets}\label{sec:Results}
To assess the methods presented above, we reproduce the parameter estimation results for the NANOGrav 15-year stochastic analysis \cite{NG15}. That is, we model intrinsic pulsar noise and a stochastic gravitational wave background across 67 pulsars observed over 15 years \cite{NG15_timing_of_68_psrs}. The white noise parameters are estimated and fixed before the parameter estimation and linear deviations to the timing model are analytically marginalized. The intrinsic pulsar red noise is modeled with a power law Eq.~(\ref{eq:power_law}) and uses $N_f=30$ frequency bins. The GWB is modeled using a power law with $N_f=14$ frequency bins, and the Hellings-Downs inter-pulsar correlation pattern Eq.~(\ref{eq:HD}) is imposed.

To test the capabilities of the standardized transformation generalized to include deterministic contributions, we \textit{simulate} a dataset consistent with the models presented in Sec.~\ref{sec:signal_comps}, including a continuous wave signal. The simulated dataset is designed to be similar to that of the NANOGrav 15-year analysis. We simulate radio pulse observations from fake pulsars at the same sky location, distance, and observational period of the NANOGrav 15-year analysis. The timing residuals are populated with simulated white noise, intrinsic pulsar red noise, a stochastic GWB signal, and a CW signal. The white noise injection is consistent with the statistics of the NANOGrav 15-year single pulsar analyses, and we inject intrinsic red noise in every pulsar according to a power law with $N_f=30$ frequency bins. A stochastic Hellings-Downs correlated GWB obeying a power law with $\log_{10}A_\text{GWB} = -14.6$ and $\gamma_\text{GWB} = 13/3$ is also injected in $N_f = 30$ frequency bins. A CW signal consistent with the model presented in \citet{CWsPTA, EllisCW, EllisCWs2} is injected with parameters $\log_{10}\mathcal{M}/[\text{M}_\odot] = 9.5$, $f_\text{CW} = 4\;\text{nHz}$, $\cos\iota = 0.6$, $\psi=0.9$, $\log_{10}h = -14.05$, $\cos\theta=0.04$, $\phi=4.06$, and $\Phi_0=1.2$. These parameters correspond to chirp mass, frequency, inclination, polarization, amplitude, polar sky location, azimuthal sky location, and phase, respectively. The injected CW signal has a matched filter signal-to-noise ratio (SNR) of 9.4 for this simulated dataset. In the posterior evaluation, the CW is represented in a Fourier domain with $N_f=60$ frequency bins.

\subsection{Results}
\begin{figure*}[t]
    \centering
    \includegraphics[width=0.9\textwidth]{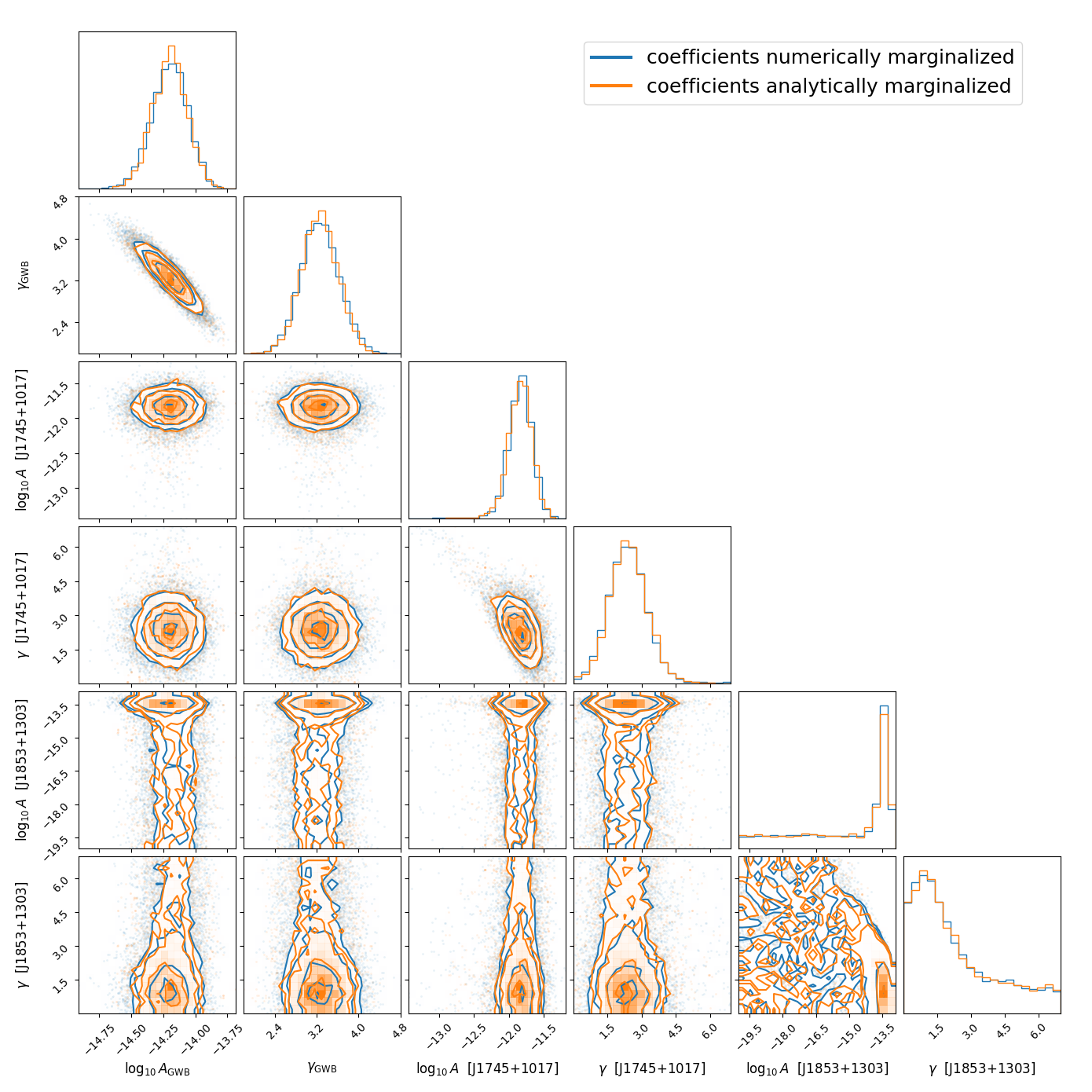}
    \caption{Posterior over the power law parameters of the stochastic GWB, J1745+1017's intrinsic noise, and J1853+1303's intrinsic noise in the NANOGrav 15-year dataset. The blue posterior is obtained using the standardizing transform and numerically marginalizing over the Fourier coefficients as presented in this paper. The orange posterior is obtained using the standard PTA analysis software \texttt{ENTERPRISE} \cite{Enterprise}, in which the Fourier coefficients are analytically marginalized from the model. The contours of the two-dimensional histograms correspond to the $(0.5, 1, 1.5, 2)-\sigma$ credible intervals.}
    \label{fig:gwb_rn_corner}
\end{figure*}

\begin{figure*}[t]
    \centering
    \includegraphics[width=0.9\textwidth]{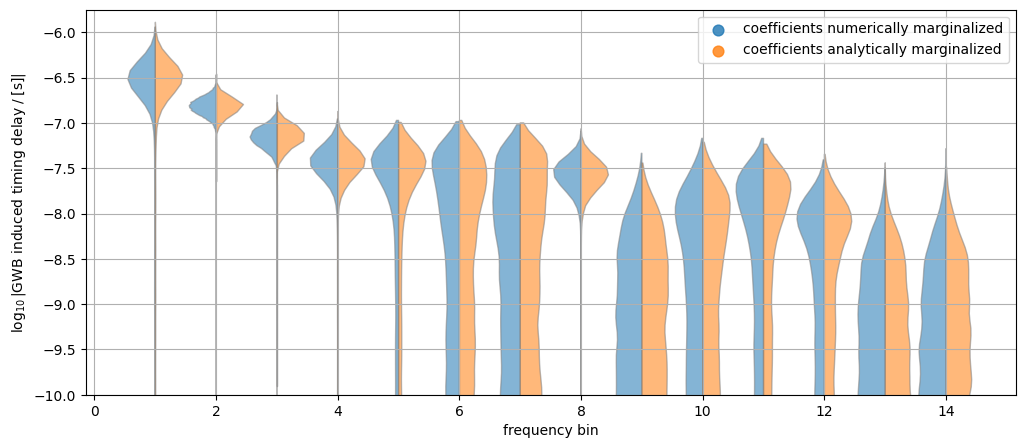}
    \caption{Free spectral analysis of the stochastic GWB in the NANOGrav 15-year dataset. The violins show the timing delays induced by the background per frequency bin. The blue violins are obtained using the standardizing transform and numerically marginalizing over the Fourier coefficients as presented in this paper. The orange violins are obtained using the standard PTA analysis software \texttt{ENTERPRISE} \citep{Enterprise}, in which the Fourier coefficients are analytically marginalized from the model.}
    \label{fig:violin}
\end{figure*}

The power law recovery of the stochastic GWB and intrinsic pulsar RN for the NANOGrav 15-year dataset are shown in Fig.~\ref{fig:gwb_rn_corner}. The posterior recovered using the methods presented in this paper is consistent with that of standard methods and results presented in \cite{NG15}. A dataset this large typically requires at least several hours to obtain a sufficient number of independent samples. On an NVIDIA GeForce RTX 3090 GPU, our analysis takes approximately 15 minutes to resolve the posterior shown in Fig~\ref{fig:gwb_rn_corner}. More precisely, the standard analytically marginalized posterior achieves $\sim 0.12$ effective samples per second while our numerically marginalized posterior under the standardizing transformation achieves $\sim 1.53$ effective samples per second, resulting in over an order of magnitude speed-up. For a fair comparison, the analytically marginalized posterior was implemented in \texttt{JAX}, used single precision, ran on the same GPU-configuration, and sampled with NUTS exactly as we do for the numerically marginalized posterior. We conclude the numerical marginalization and standardizing transformation approach is over an order of magnitude faster than the analytically marginalized analysis. The same dataset is also analyzed using a free spectral model to represent the frequency content of the stochastic GWB. The magnitude of the timing delays induced by the GWB per frequency bin is shown in Fig.~\ref{fig:violin}, and is consistent with the standard results of \cite{NG15}.

\begin{figure*}
    \centering
    \includegraphics[width=0.95\linewidth]{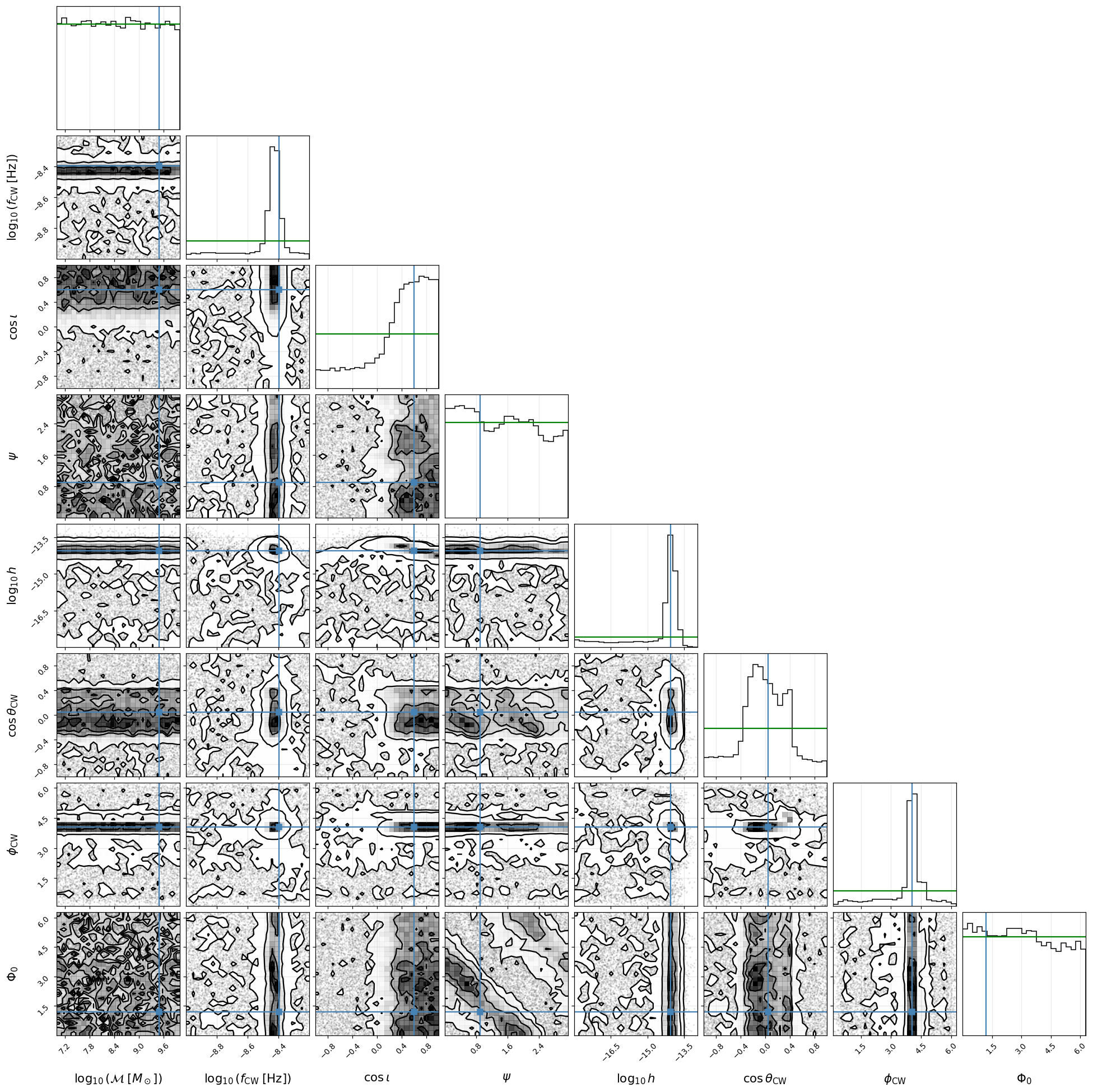}
    \caption{Posterior over the CW signal's source parameters in the simulated dataset. The CW signal has a matched-filter $\text{SNR}\approx 9.4$. Blue lines indicate injected parameter values. The green distribution along the diagonal is the prior over the CW source parameters. The contours of the two-dimensional histograms correspond to the $(0.5, 1, 1.5, 2)-\sigma$ credible intervals.}
    \label{fig:cw_corner}
\end{figure*}

The simulated dataset is analyzed under a model consistent with the injection specified above. The recovery of the CW source parameters for this simulated dataset are shown in Fig.~\ref{fig:cw_corner}. The injected parameter values lie within the posterior distribution. It's worth emphasizing this analysis jointly models a HD-correlated GWB, intrinsic pulsar RN, and a CW simultaneously in a relatively large PTA. Such analyses were extremely computationally intensive, typically requiring many hours of computation time or various approximations before this work. For example, the \texttt{QuickCW} software \citep{QuickCW} neglects inter-pulsar correlations and models a CURN GWB and a CW simultaneously. If a joint HD-correlated GWB and CW analysis was desired, the samples from \texttt{QuickCW} could be reweighted to include inter-pulsar correlations. However, very few independent samples survive the reweighting process requiring lengthy runs to build up the initial sample set \citep{NG15_CW}. Using the methods presented above, the sampling chain for the joint HD-correlated GWB + CW + RN model converged in less than 30 minutes on an NVIDIA GeForce RTX 3090.

The posterior over the CW source parameters in the simulated dataset, Fig.~\ref{fig:cw_corner}, exhibits some multi-modality, particularly in the polarization and phase parameters. This is expected as the CW model exhibits an exact discrete symmetry under the transformation $\{\psi, \Phi_0\}\rightarrow\{\psi+\pi/2,\Phi_0 + \pi\}$. A known weakness of the HMC and NUTS schemes is the tendency for sampling chains to get stuck on local maxima and fail to explore the typical set of the posterior. We are able to verify, for this example, that our chain explores the entire posterior from Fig.~\ref{fig:cw_corner} given that almost the entire prior volume is explored. The only region of parameter space with no significant probability mass is $\log_{10}h \gtrsim -13.5$, which serves as an upper bound as there is no evidence for such a loud CW signal. In particular, our simulation (evidenced by the one-dimensional marginal distribution over $\log_{10}h$) is in the \textit{weakly-detectable} regime where the sampling chain mixes between states which either latch onto the CW signal or explore the prior distribution, restricted by the upper bound $\log_{10}h \lesssim-13.5$. We are therefore confident our sampling strategy will perform well in CW all-sky searches in real datasets, primarily because there has been little evidence thus far for a signal from an individual SMBHB, and current searches are exploring similar weakly-detectable regimes.

On the other hand, we might choose to model deterministic signals which exhibit stronger multi-modality, with sharp peaks significantly separated in parameter space. Searches for GW bursts with memory~\cite{bursts}, targeted searches for individual binaries~\cite{targeted_search}, and CWs of greater SNR will induce such multi-modal and non-Gaussian posterior geometries. In such cases, it will be necessary to improve the proposal distribution used in the sampling scheme, as NUTS will tend to get stuck on local maxima. A potential solution is to mix parallel tempering~\cite{PTMCMC} (a sampling scheme designed to jump between posterior modes) with NUTS proposals. The tempered posterior and standardizing transformation necessary for parallel tempering is derived in Appendix~\ref{app:temp_like} as a first step toward more robust determinstic signal sampling, but is not implemented here.

\begin{figure}
    \centering
    \includegraphics[width=0.9\linewidth]{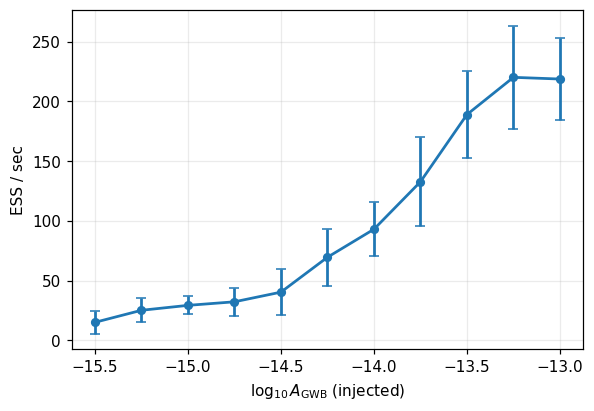}
    \caption{The effective sample size per second (ESS/sec) achieved with NUTS and the (CURN-approximated) standardizing transformation. For every data point, we inject five realizations of the background and sample each of those five times with a different random initialization seed. The error bars are $1\sigma$ variations in the ESS/sec. Note this is not a \textit{global} sampling rate, and the ESS/sec will depend on the particular models and datasets analyzed. We show how the posterior evaluation under the standardizing transformation scales with the number of pulsars and frequency bins in Sec.~\ref{subsec:batching}.}
    \label{fig:ESS/sec}
\end{figure}

At face value, it may be concerning that the standardizing transformation only approximately de-correlates the parameter space. One might envision a high-resolution PTA, with a booming loud background in which HD-correlations are constrained with great precision. Does the CURN-approximated standardizing transformation lose effectiveness for such systems? The answer is no. Recall the transformation uses statistical moments, $\mathbf{\hat{a}}_\text{CURN} = \boldsymbol{\Sigma}_\text{CURN}\mathbf{F}^\text{T}\mathbf{\tilde{N}}^{-1}\boldsymbol{\delta t}$ and $\boldsymbol{\Sigma}_\text{CURN}^{-1}=\mathbf{F}^\text{T}\mathbf{\tilde{N}}^{-1}\mathbf{F} + \boldsymbol{\phi}_\text{CURN}^{-1}$. The covariance contains contributions from the likelihood ($\mathbf{F}^\text{T}\mathbf{\tilde{N}}^{-1}\mathbf{F}$) and prior ($\boldsymbol{\phi}^{-1}_\text{CURN}$). In the high SNR limit, the amplitude of the background, $A_\text{GWB}$, increases such that $\boldsymbol{\Sigma}^{-1}_\text{CURN} \rightarrow \mathbf{F}^\text{T}\mathbf{\tilde{N}}^{-1}\mathbf{F}$ and $\hat{\mathbf{a}}_\text{CURN}\rightarrow (\mathbf{F}^\text{T}\mathbf{\tilde{N}}^{-1}\mathbf{F})^{-1}\mathbf{F}^\text{T}\tilde{\mathbf{N}}^{-1}\boldsymbol{\delta t}$. The standardizing transformation becomes likelihood-dominated, mirroring the posterior. In other words, the CURN-approximation only modifies the prior component of the transformation which is overwhelmed by the likelihood term as the strength of the signal increases. This limit is expected: in the data-dominated regime, the inverse covariance converges to the Fisher information matrix, and the MAP Fourier coefficients reduce to their maximum likelihood values which is the least squares solution. In Fig.~\ref{fig:ESS/sec}, we show the standardizing transformation does not hinder sampling empirically by measuring the effective sample size per second (ESS/sec) across a series of background injections of increasing strength. As the strength of the background increases the (approximated) prior term in transformation loses significance, and the ESS/sec increases, \textit{improving} the sampling.

\subsection{Discussion and future work}
Historically, PTA data analysis has been computationally expensive. The methods presented in this paper dramatically improve efficiency, yielding more than an order of magnitude speed-up for analysis of realistic datasets. Rather than analytically marginalizing over the set of Fourier coefficients which represent stochastic red signals and noise, they are sampled (i.e. numerically marginalized) with a hyper-efficient posterior formulation as presented in \cite{Lentati}. While hyper-efficient, the posterior is high-dimensional and exhibits a complicated funnel-like geometry from which it is generally difficult to sample. We reparameterize the posterior with a standardizing coordinate transform so the Fourier coefficients are approximately described by a standard normal distribution. The high-dimensional posterior can then be efficiently sampled using HMC with the NUTS when run on a GPU.

The standardizing transform for PTAs is generalized to include contributions from arbitrary deterministic signals. The recovery of a particular simulated CW signal is shown above, but future work needs to improve the sampling of deterministic signal parameters which may induce complicated posterior geometries. The posteriors under deterministic models are often multi-modal, significantly non-Gaussian, and troublesome for standard HMC samplers. Deterministic signal analysis may be improved by mixing other jump proposals, such as parallel tempering, with classic HMC techniques, see Appendix~\ref{app:temp_like}.

\begin{acknowledgments}
\textit{Software}. The PTA analyses presented here were conducted with the \texttt{prometheus} Python package \citep{Prometheus}, which implements the standardizing transformation methods above. These methods are currently being added to the \texttt{discovery} \citep{Discovery} and \texttt{Atlas} (in preparation) packages as well for next-generation PTA analyses. Within \texttt{prometheus}, the posterior density and standardizing transform are implemented in \texttt{JAX} \citep{JAX}. The sampling is performed with \texttt{NumPyro} \citep{NumPyro, NumPyro2}. Figures are created with \texttt{matplotlib} \citep{matplotlib} and \texttt{corner} \citep{corner.py}. The code and simulated datasets to reproduce figures presented in this work are publicly available at \cite{code_for_figures}.

\textit{Data}. Simulated data is generated with the \texttt{NumPy} package \citep{NumPy}. The North American Nanohertz Observatory for Gravitational Waves (NANOGrav) 15-year Data Set \cite{NG15_data} is analyzed using the methods presented in this paper to reproduce existing results \cite{NG15}.

\textit{Funding}. The authors are members of the NANOGrav collaboration supported by National Science Foundation Physics Frontiers Center grant No. 2020265.
\end{acknowledgments}

\appendix

\section{Hamiltonian Monte Carlo and No U-Turn Sampling}\label{app:HMC}
Hamiltonian Monte Carlo (HMC), originally known as \textit{Hybrid Monte Carlo}, was first developed for calculations in lattice quantum chromodynamics \cite{QCD_HMC}. It was later popularized for applied statistics in \cite{Neal1996, Neal2011}. \cite{BetancourtHMC} illustrated its robustness by setting it in the language of differential geometry. The sampling above is performed with HMC and its No U-Turn Sampler (NUTS) extension and implemented in \texttt{NumPyro} \cite{NumPyro, NumPyro2}. We summarize HMC and NUTS in this appendix, following the treatment of \cite{concept_HMC} to which the reader is directed for a more thorough discussion.

The aim of Markov Chain Monte Carlo (MCMC) is to sample from a target distribution, $p(\mathbf{x})$, under some parameterization $\mathbf{x}\in\mathcal{X}$, where $\mathcal{X}$ is the target sample space and $\text{dim}(\mathbf{x}) = d$. MCMC is performed by constructing a Markov Chain whose equilibrium distribution is the target distribution itself. Once sufficiently many samples have been drawn, they may be binned into a histogram to reconstruct the target distribution, or approximate expectations using Monte Carlo integration,
\begin{equation}
    \mathbb{E}[f] = \int_\mathcal{X}f(\mathbf{x})\,p(\mathbf{x})\,d\mathbf{x} \approx \frac{1}{N}\sum_i^N f(\mathbf{x}^{(i)})
\end{equation}
where $\mathbf{x}^{(i)}$ denotes the $i^\text{th}$ of $N$ total random samples from the target distribution.

The standard algorithm to produce random samples from a target distribution is the Metropolis-Hastings (MH) MCMC algorithm, \cite{Metropolis, Metropolis-Hastings}. The MH algorithm initializes a random sample $\mathbf{x}^{(0)}\in\mathcal{X}$, and after $i$ iterations, the chain may transition from state $\mathbf{x}^{(i)}$ to $\mathbf{y}$. The transition is proposed by drawing from the proposal distribution $g(\mathbf{y}|\mathbf{x}^{(i)})$, and accepted with transition probability
\begin{equation}
    \alpha = \text{min}\bigg\{1,\;\;\frac{p(\mathbf{y})\cdot g(\mathbf{x}^{(i)}|\mathbf{y})}{p(\mathbf{x}^{(i)})\cdot g(\mathbf{y}|\mathbf{x}^{(i)})}\bigg\}\,.
\end{equation}
If the proposal is accepted, the next sample is set to the proposal, $\mathbf{x}^{(i+1)}=\mathbf{y}$. If the proposal is rejected, the chain remains at the same sample, $\mathbf{x}^{(i+1)}=\mathbf{x}^{(i)}$. The MH algorithm is repeated for the desired number of samples. If the proposal density is sufficiently well-tailored, then the algorithm is guaranteed to produce samples which converge to the target distribution. Samples obtained via MCMC are correlated with one another, by construction, and the chain of samples usually needs to be \textit{thinned} until only statistically independent samples remain.

HMC operates in a similar fashion to MH, but the proposals are not drawn from some proposal distribution and implemented in random walk fashion. Instead, the proposals are generated by integrating Hamilton's equations of motion, which dictate the chain's evolution deterministically through a phase space (inspired by Hamilton's formulation of classical mechanics). The parameters of the target density are extended to include a set of canonical momenta, $\mathbf{x}\rightarrow (\mathbf{x}, \mathbf{k})\in\mathcal{P}$, where $\mathbf{x}$ assume the role of generalized coordinates, $\mathbf{k}$ canonical momenta, $\text{dim}(\mathbf{k})=d$, and $\mathcal{P}$ denotes the phase space. The target density is similarly lifted to the \textit{canonical distribution} with support on the entire $2d$-dimensional phase space,
\begin{equation}
    p(\mathbf{x},\mathbf{k}) = p(\mathbf{k}|\mathbf{x})\cdot p(\mathbf{x})\,.
\end{equation}
We recover the target distribution, $p(\mathbf{x})$, if the canonical momenta are marginalized over.

The Hamiltonian governing dynamics on the phase space is defined as
\begin{align}\label{eq:Hamiltonian}
    H(\mathbf{x}, \mathbf{k}) &\equiv -\ln p(\mathbf{x}, \mathbf{k}) \nonumber \\
    &= -\ln p(\mathbf{k}|\mathbf{x}) -\ln p(\mathbf{x})\,.
\end{align}
The two terms in Eq.~(\ref{eq:Hamiltonian}) are the so called \textit{kinetic} and \textit{potential energy}, respectively. Trajectories through this phase space are described by Hamilton's equations of motion, which are $2d$ coupled first-order ordinary differential equations,
\begin{align}\label{eq:Hamilton's_eq}
    \dot{\mathbf{x}}^i &= \frac{\partial H}{\partial\mathbf{k}_i} \nonumber \\
    \dot{\mathbf{k}}_i &= -\frac{\partial H}{\partial\mathbf{x}^i}\,,
\end{align}
where $i\in\{1,2,\dots,d\}$ indexes the components of the generalized coordinates and canonical momenta. To complete the probabilistic structure on phase space, we must choose a form for the kinetic energy, $p(\mathbf{k}|\mathbf{x})$. Rather than searching over infinite possibilities for the optimal kinetic energy, we'll restrict ourselves to \textit{Euclidean-Gaussian kinetic energies} in which the distance between two configurations with coordinates $\mathbf{x}$ and $\mathbf{x}'$ is
\begin{equation}
    \Delta(\mathbf{x}, \mathbf{x}') = \mathbf{g}_{ij}\,(\mathbf{x}-\mathbf{x}')^i\,(\mathbf{x}-\mathbf{x}')^j\,,
\end{equation}
where $i,j\in\{1,2,\dots,d\}$, $\mathbf{g}_{ij}$ are the components of the Euclidean metric tensor, and the Einstein summation convention is used. In order to preserve volumes in phase space under coordinate transformations, the canonical momenta must use the inverse transformation to that of the generalized coordinates when performing reparameterizations. The distance between momenta $\mathbf{k}$ and $\mathbf{k}'$ is
\begin{equation}
    \Delta(\mathbf{k}, \mathbf{k}') = \mathbf{g}^{ij}\,(\mathbf{k}-\mathbf{k}')_i\,(\mathbf{k}-\mathbf{k}')_j\,,
\end{equation}
where $\mathbf{g}^{ij}$ are the components of the \textit{inverse} Euclidean metric. The metric allows the construction of probability densities over the momenta, and the Euclidean-Gaussian (zero mean) kinetic energy is
\begin{equation}\label{eq:Gaussian_momentum}
    p(\mathbf{k}|\mathbf{x}) = \frac{1}{\sqrt{\text{det}(2\pi\mathbf{M})}}\,\text{exp}\bigg[-\frac{1}{2}\mathbf{k}^\text{T}\,\mathbf{M}^{-1}\,\mathbf{k}\bigg]\,,
\end{equation}
where we've adopted the physics notation of a \textit{mass matrix}, $\mathbf{M}$, to replace the inverse metric tensor. We have not yet chosen a form for the Euclidean metric (or mass matrix), and an optimal choice for arbitrary systems has not yet been derived. However, acting as a covariance matrix, the mass matrix has the ability to de-correlate the parameters of phase space. We may de-correlate the parameters of the target density, $\mathbf{x}$, by estimating their covariance, and because they transform in opposite fashion, the inverse covariance should be used to define the mass matrix,
\begin{equation}
    \mathbf{M}^{-1} = \text{cov}(\mathbf{x}, \mathbf{x})\,.
\end{equation}
Geometrically, this choice distributes the level sets of the Hamiltonian in phase space so that the sampling is easier. In practice, the covariance is often estimated empirically with a set of warm-up samples. No global transformation using the covariance is going to perfectly de-correlate the target space for arbitrary problems, unless they are perfectly Gaussian. To improve sampling further for systems with significant non-Gaussian features, it is possible to extend beyond Euclidean geometries and use a \textit{Riemannian-Gaussian kinetic energy}. For such systems the covariance is computed locally, $\boldsymbol{\Sigma} = \boldsymbol{\Sigma}(\mathbf{x})$, using say a Fisher information matrix approach, $\boldsymbol{\Sigma}^{-1}(\mathbf{x})\approx-\partial_\mathbf{x}\partial_\mathbf{x}\ln p(\mathbf{x})$. Then the kinetic energy assumes a multivariate normal distribution, whose covariance is a position-dependent field over the target space,
\begin{equation}
    \mathbf{k}|\mathbf{x}\sim\mathcal{N}(\mathbf{0}, \boldsymbol{\Sigma}(\mathbf{x}))\,.
\end{equation}

Sampling with HMC is then performed as follows. At the $i^\text{th}$ iteration we have sample $\mathbf{x}^{(i)}$. To generate the $(i + 1)^\text{th}$ sample, we first draw a set of momenta, $\mathbf{k}^{(i)}$, from the kinetic energy distribution, Eq.~(\ref{eq:Gaussian_momentum}). $(\mathbf{x}^{(i)}, \mathbf{k}^{(i)})$ serve as the initial conditions to Hamilton's equations of motion, Eq.~(\ref{eq:Hamilton's_eq}), which are integrated so the chain evolves through phase space. After some time, $\Delta t$, the integration is terminated and the chain's state is $(\mathbf{x}_{\Delta t}, \mathbf{k}_{\Delta t})$. This state is the proposal for the next sample and is accepted with probability
\begin{equation}\label{eq:HMC_acceptance}
    \alpha = \text{min}\bigg\{1, \frac{\text{exp}[-H(\mathbf{x}_{\Delta t}, \mathbf{k}_{\Delta t})]}{\text{exp}[-H(\mathbf{x}^{(i)}, \mathbf{k}^{(i)})]}\bigg\}\,.
\end{equation}

If the integration is exact, then the chain evolves along a level set of the Hamiltonian and the proposal is always accepted. In practice, the integration is performed numerically with a leapfrog integrator, see Sec. 5.1 of \cite{concept_HMC}. Leapfrog integration is symplectic, conserving volume in phase space, and the trajectories never stray far from level set at which they were initialized. Symplectic integrators introduce errors that are tangent to a nearby energy surface, while non-symplectic (such as primitive Euler or Runge–Kutta) integrators introduce errors with a systematic component normal to energy surfaces. HMC can achieve relatively high (often $\sim90\%$) acceptance rates thanks to symplectic integrators moving along level sets of the Hamiltonian. At subsequent iterations, the momenta are redrawn from Eq.~(\ref{eq:Gaussian_momentum}), and the chain is integrated along a new level set. If the mass matrix accurately describes the distribution of level sets, the typical set of the target distribution is quickly explored after relatively few draws from the kinetic energy distribution.

The final piece is to decide how long to integrate Hamilton's equations at each iteration. If $\Delta t$ is too short, the chain is highly correlated and resembles random walk. If $\Delta t$ is too long, significant computation time is spent integrating Hamilton's equations for every sample. The No U-turn termination condition, \cite{NUTS}, provides an empirical method to decide integration time. It integrates trajectories through the phase space until a ``U-turn" is detected. This encourages trajectories long enough to reduce auto-correlations, but not so long they double back on themselves, wasting integration time in a previously covered region of phase space.

\section{The standardizing transform for tempered likelihoods}\label{app:temp_like}
While HMC is a highly efficient sampling scheme in many systems, the chain will get stuck on local maxima if the posterior exhibits multi-modality. For multi-modal target densities a parallel tempered sampling scheme~\citep{PTMCMC} may be appropriate in which the log-likelihood is scaled by a temperature $\mathcal{T}$ to aid chain exploration. Below we generalize the standardizing transform to allow temperature scaling of the chain.

In a tempering scheme, the likelihood is raised to the power of $\beta = 1/\mathcal{T}$ so the tempered posterior may be written
\begin{equation}
    p_\beta(\mathbf{a},\boldsymbol{\eta}|\boldsymbol{\delta t}) \propto p(\boldsymbol{\delta t}|\mathbf{a})^\beta \cdot p(\mathbf{a}|\boldsymbol{\eta}) \cdot p(\boldsymbol{\eta})
\end{equation}
such that for large temperatures the likelihood is suppressed and we recover a distribution which more closely resembles the prior. In the infinite temperature limit the tempered posterior converges to the prior. Tempering the posterior, Eq.~(\ref{eq:raw_posterior}), we have
\begin{widetext}
\begin{equation}\label{eq:tempered_post}
    p_\beta(\mathbf{a}, \boldsymbol{\eta}|\boldsymbol{\delta t}) \propto \frac{p(\boldsymbol{\eta})}{\text{det}(2\pi\tilde{\mathbf{N})}^{\beta/2}}\,\text{exp}\bigg[-\frac{\beta}{2}\big(\boldsymbol{\delta t} - \mathbf{F}\mathbf{a}\big)^\text{T}\,\mathbf{\tilde{N}}^{-1}\,\big(\boldsymbol{\delta t} - \mathbf{F}\mathbf{a}\big)\bigg]\times\frac{1}{\sqrt{\text{det}(2\pi\boldsymbol{\phi})}}\text{exp}\bigg[-\frac{1}{2}\mathbf{a}^\text{T}\,\boldsymbol{\phi}^{-1}\,\mathbf{a}\bigg]\,.
\end{equation}
\end{widetext}
Completing the square, the posterior can be expressed as
\begin{widetext}
\begin{equation}\label{eq:mv_normal_post_temp}
    p_\beta(\mathbf{a},\boldsymbol{\eta}|\boldsymbol{\delta t})\propto \frac{p(\boldsymbol{\eta})}{\sqrt{\text{det}(2\pi\boldsymbol{\phi})}}\,\text{exp}\bigg[-\frac{1}{2}\big(\mathbf{a} - \mathbf{\hat{a}}_\beta\big)^\text{T}\,\boldsymbol{\Sigma}_\beta^{-1}\,\big(\mathbf{a} - \mathbf{\hat{a}}_\beta\big) + \frac{1}{2}\mathbf{\hat{a}}_\beta^\text{T}\,\boldsymbol{\Sigma}_\beta^{-1}\,\mathbf{\hat{a}}_\beta\bigg]
\end{equation}
\end{widetext}
where the estimated mean and covariance of the Fourier coefficients is $\hat{\mathbf{a}}_\beta = \beta\cdot\boldsymbol{\Sigma}_\beta\mathbf{F}^\text{T}\tilde{\mathbf{N}}^{-1}\boldsymbol{\delta t}$ and $\boldsymbol{\Sigma}_\beta = \big(\beta\cdot\mathbf{F}^\text{T}\tilde{\mathbf{N}}^{-1}\mathbf{F} + \boldsymbol{\phi}^{-1}\big)^{-1}$, respectively. The mean and covariance are not scaled trivially by the temperature as they involve likelihood and prior contributions, and only terms in the likelihood are scaled by temperature. The tempered standardizing transform is
\begin{equation}
    (\mathbf{a}, \boldsymbol{\eta}) = T_\beta^{-1}(\mathbf{z}, \boldsymbol{\eta}) = (\hat{\mathbf{a}}_\beta + \mathbf{L}_\beta\,\mathbf{z}, \boldsymbol{\eta})
\end{equation}
where $\boldsymbol{\Sigma}_\beta = \mathbf{L}_\beta\mathbf{L}_\beta^\text{T}$ is the Cholesky decomposition of tempered covariance matrix. While the sampling of a tempered posterior under an un-tempered standardizing transform is valid, the transform will consistently map the Fourier coefficients into a narrower region of parameter space relative to the tempered posterior volume, and the chain will take longer to converge. Beyond multi-modality tempered posteriors are useful in the calculation of evidences and Bayes factors via thermodynamic integration \cite{thermo_int}.

\section{The standardizing transform for split coefficients.}\label{app:split_coeffs}

Rather than combining the intrinsic pulsar red noise and gravitational wave background into one Gaussian process as in Eq.~(\ref{eq:prior_cov}), we may wish to treat them as independent stochastic processes \footnote{This is helpful when we wish to model inter-frequency correlations (see Sec.~\ref{subsec:inter-freq}) or non-Gaussian features (see Appendix~\ref{app:non-gauss}).}. This choice necessitates using two distinct sets of Fourier coefficients in the posterior: one to represent pulsar noise and one to represent the background.

If we use separate sets of Fourier coefficients to describe the intrinsic pulsar red noise and the stochastic gravitational wave background, the posterior Eq.~(\ref{eq:raw_posterior}) is modified
\begin{widetext}
\begin{align}\label{eq:separated_post}
    p(\mathbf{a}_\text{GWB}, \mathbf{a}_\text{RN},\boldsymbol{\eta}|\boldsymbol{\delta t})\propto &\frac{p(\boldsymbol{\eta})}{\sqrt{\text{det}(2\pi\tilde{\mathbf{N}})}}\,\text{exp}\bigg[-\frac{1}{2}\big(\boldsymbol{\delta t} - \mathbf{F}\mathbf{a}_\text{GWB}-\mathbf{F}\mathbf{a}_\text{RN}\big)^\text{T}\,\mathbf{\tilde{N}}^{-1}\,\big(\boldsymbol{\delta t} - \mathbf{F}\mathbf{a}_\text{GWB}-\mathbf{F}\mathbf{a}_\text{RN}\big)\bigg]  \\ \nonumber
    &\times\frac{1}{\sqrt{\text{det}(2\pi\boldsymbol{\phi}_\text{GWB})}}\text{exp}\bigg[-\frac{1}{2}\mathbf{a}_\text{GWB}^\text{T}\,\boldsymbol{\phi}_\text{GWB}^{-1}\,\mathbf{a}_\text{GWB}\bigg] \\ \nonumber
&\times\frac{1}{\sqrt{\text{det}(2\pi\boldsymbol{\phi}_\text{RN})}}\text{exp}\bigg[-\frac{1}{2}\mathbf{a}_\text{RN}^\text{T}\,\boldsymbol{\phi}_\text{RN}^{-1}\,\mathbf{a}_\text{RN}\bigg]\,,
\end{align}
\end{widetext}
where the subscripts ``GWB" and ``RN" represent the background and intrinsic noise respectively. Both sets of coefficients are defined with respect to the same basis $\mathbf{F}$, and a simple replacement $\mathbf{F}\rightarrow\mathbf{F}_\text{RN},\mathbf{F}_\text{GWB}$ is used to represent the processes with respect to distinct Fourier bases. The MAP Fourier coefficients, are defined by the extremum condition,
\begin{equation}
    \begin{pmatrix}
        \partial_{\mathbf{a}_\text{GWB}}\ln p(\mathbf{a}_\text{GWB}, \mathbf{a}_\text{RN}, \boldsymbol{\eta}|\boldsymbol{\delta t})\big\vert_{\hat{\mathbf{a}}_\text{GWB}, \hat{\mathbf{a}}_\text{RN}} \\
        \partial_{\mathbf{a}_\text{GWB}}\ln p(\mathbf{a}_\text{GWB}, \mathbf{a}_\text{RN}, \boldsymbol{\eta}|\boldsymbol{\delta t})\big\vert_{\hat{\mathbf{a}}_\text{GWB}, \hat{\mathbf{a}}_\text{RN}}
        
    \end{pmatrix} = \begin{pmatrix}
    0 \\
    0
    \end{pmatrix}
\end{equation}
which yields the linear system
\begin{equation}\label{eq:split_map_a}
    \begin{pmatrix}
        \boldsymbol{\Sigma}_\text{GWB}^{-1} & \mathbf{F}^\text{T}\tilde{\mathbf{N}}^{-1}\mathbf{F} \\
        \mathbf{F}^\text{T}\tilde{\mathbf{N}}^{-1}\mathbf{F} & \boldsymbol{\Sigma}_\text{RN}^{-1}
    \end{pmatrix}\begin{pmatrix}
        \hat{\mathbf{a}}_\text{GWB} \\
        \hat{\mathbf{a}}_\text{RN}
    \end{pmatrix} = \begin{pmatrix}
        \mathbf{F}^\text{T}\tilde{\mathbf{N}}^{-1}\boldsymbol{\delta t} \\
        \mathbf{F}^\text{T}\tilde{\mathbf{N}}^{-1}\boldsymbol{\delta t}
    \end{pmatrix}\,,
\end{equation}
where the covariances, estimated from the Hessian of the log-posterior, are $\boldsymbol{\Sigma}_\text{GWB} = \big(\mathbf{F}^\text{T}\tilde{\mathbf{N}}^{-1}\mathbf{F} + \boldsymbol{\phi}_\text{GWB}^{-1}\big)^{-1}$ and $\boldsymbol{\Sigma}_\text{RN} = \big(\mathbf{F}^\text{T}\tilde{\mathbf{N}}^{-1}\mathbf{F} + \boldsymbol{\phi}_\text{RN}^{-1}\big)^{-1}$. While Eq.~(\ref{eq:split_map_a}) may be solved analytically, it is more computationally efficient to solve it numerically using a Cholesky decomposition. The standardizing transformation is then modified
\begin{widetext}
\begin{equation}\label{eq:split_standardizing_transform}
    (\mathbf{a}_\text{GWB}, \mathbf{a}_\text{RN}, \boldsymbol{\eta}) = T^{-1}(\mathbf{z}_\text{GWB}, \mathbf{z}_\text{RN}, \boldsymbol{\eta}) = (\hat{\mathbf{a}}_\text{GWB} + \mathbf{L}_\text{GWB}\mathbf{z}_\text{GWB}, \hat{\mathbf{a}}_\text{RN} + \mathbf{L}_\text{RN}\mathbf{z}_\text{RN}, \boldsymbol{\eta})
\end{equation}
\end{widetext}
where $\mathbf{L}_\text{GWB/RN}$ is the Cholesky decomposition of the covariance matrix $\boldsymbol{\Sigma}_\text{GWB/RN}$. As above, we approximate the covariance of background by neglecting inter-pulsar correlations in the transformation. By definition, the intrinsic pulsar red noise is independent across pulsars and its covariance matrix is already diagonal in the pulsar basis. Notice the covariance of the sets of Fourier coefficients is determined independently from the hyper-parameters of the respective stochastic process while the MAP GWB and RN coefficients are coupled.

\section{Non-Gaussian features}\label{app:non-gauss}
It is well known in the PTA community that stochastic astrophysical signals are non-Gaussian. Namely the stochastic gravitational wave background, if realized via a finite population of supermassive black hole binaries, will exhibit statistical moments beyond second-order, \cite{Higher_moments, Higher_moments2, Higher_moments3, GWB_dist1, GWB_dist2, GWB_dist3}. However for modeling efficiency, such signals are often approximated with a Gaussian distribution (as we have done above Eq.~(\ref{eq:a_prior})). As the sensitivity of PTAs improve non-Gaussian features may become resolvable, and we'll need analysis pipelines capable of modeling higher-order statistical moments.

Standard analyses, in which the Fourier coefficients are analytically marginalized from the model, will struggle to model such non-Gaussian features. The analytic marginalization is only possible because a conjugate prior is chosen for the Fourier coefficients, Eq.~(\ref{eq:a_prior}), so that both the posterior and prior are a multivariate normal distribution in the Fourier coefficients, Eq.~(\ref{eq:stochastic_posterior}), for which a closed form analytic expression for the marginalizing integral is known. If a prior for the Fourier coefficients with non-Gaussian features is chosen, then analytic marginalization may not be possible. Methods have been developed to model non-Gaussian features in PTA datasets nonetheless, e.g. via particular parameterizations of non-Gaussian features \cite{nonGauss_Lentati} or Gaussian mixture models \cite{nonGauss_Falxa}. In this appendix, we show how non-Gaussian priors on the Fourier coefficients may be implemented in tandem with the methods presented above and demonstrate how such non-Gaussian features can be resolved efficiently by analyzing a simulated dataset.

As we sample the Fourier coefficients jointly, our methods don't require an analytic marginalization and we are free to choose arbitrary priors for the coefficients. Let $q(\mathbf{a}|\boldsymbol{\eta})$ be an arbitrary non-Gaussian prior on the coefficients, conditioned on the spectral hyper-parameters $\boldsymbol{\eta}$. Then the PTA posterior is proportional to
\begin{widetext}
\begin{align}\label{eq:non_gauss_post}
    p(\mathbf{a},\boldsymbol{\eta}|\boldsymbol{\delta t}) &\propto p(\boldsymbol{\delta t}|\mathbf{a}) \cdot q(\mathbf{a}|\boldsymbol{\eta}) \cdot p(\boldsymbol{\eta}) \nonumber \\
    &\propto \frac{q(\mathbf{a}|\boldsymbol{\eta}) \cdot p(\boldsymbol{\eta})}{\sqrt{\text{det}(2\pi\tilde{\mathbf{N}})}}\,\text{exp}\bigg[-\frac{1}{2}\big(\boldsymbol{\delta t} - \mathbf{F}\mathbf{a}\big)^\text{T}\,\mathbf{\tilde{N}}^{-1}\,\big(\boldsymbol{\delta t} - \mathbf{F}\mathbf{a}\big)\bigg] \nonumber \\
    &\propto \frac{q(\mathbf{a}|\boldsymbol{\eta}) \cdot p(\boldsymbol{\eta})}{\sqrt{\text{det}(2\pi\tilde{\mathbf{N}})}}\,\text{exp}\bigg[\boldsymbol{\delta t}^\text{T}\,\tilde{\mathbf{N}}^{-1}\,\mathbf{F}\mathbf{a} - \frac{1}{2}\mathbf{a}^\text{T}\,\mathbf{F}^\text{T}\,\tilde{\mathbf{N}}^{-1}\,\mathbf{F}\mathbf{a}\bigg]
\end{align}
\end{widetext}
where we have opted to analytically marginalize over linear deviations to the timing model for convenience. Note we may compute $\boldsymbol{\delta t}^\text{T}\tilde{\mathbf{N}}^{-1}\mathbf{F}$ and $\mathbf{F}^\text{T}\tilde{\mathbf{N}}^{-1}\mathbf{F}$ once and store for all future evaluations, so Eq.~(\ref{eq:non_gauss_post}) is as computationally efficient to evaluate as the Gaussian posterior, Eq.~(\ref{eq:stochastic_posterior}). The only additional cost may come from the evaluation of the modified prior $q(\mathbf{a}|\boldsymbol{\eta})$ with non-Gaussian moments.

Lastly, we must determine the proper standardizing transform for this posterior. While $q(\mathbf{a}|\boldsymbol{\eta})$ is non-Gaussian, we'll assume the Laplace approximation is valid in a neighborhood about the MAP solution, $\mathbf{a}=\check{\mathbf{a}}$,
\begin{equation}
    q(\mathbf{a}|\boldsymbol{\eta})\bigg\vert_{\mathbf{a}\approx\check{\mathbf{a}}}\approx \frac{1}{\sqrt{\text{det}(2\pi\boldsymbol{\varphi})}}\,\text{exp}\bigg[-\frac{1}{2}\big(\mathbf{a}-\boldsymbol{\xi}\big)^\text{T}\,\boldsymbol{\varphi}^{-1}\,\big(\mathbf{a}-\boldsymbol{\xi}\big)\bigg]
\end{equation}
where $\boldsymbol{\xi}$ and $\boldsymbol{\varphi}$ are the mean and covariance of the non-Gaussian prior under the Laplace approximation. Then we may estimate the MAP coefficients and the covariance of the posterior using the extremum condition and the Hessian of the log-posterior to find $\check{\mathbf{a}} = \boldsymbol{\Psi}\big(\mathbf{F}^\text{T}\tilde{\mathbf{N}}^{-1}\boldsymbol{\delta t} + \boldsymbol{\varphi}^{-1}\boldsymbol{\xi}\big)$ and $\boldsymbol{\Psi} = \big(\mathbf{F}^\text{T}\tilde{\mathbf{N}}^{-1}\mathbf{F} + \boldsymbol{\varphi}^{-1}\big)^{-1}$, respectively. Under the replacement $\hat{\mathbf{a}}\rightarrow\check{\mathbf{a}}$ and $\boldsymbol{\Sigma}\rightarrow\boldsymbol{\Psi}$ we may write the standardizing transform, Eq.~(\ref{eq:PTA_standard_transform}), as
\begin{equation}\label{eq:non_gauss_standard_transform}
    (\mathbf{a}, \;\boldsymbol{\eta}) = T^{-1}(\mathbf{z}, \boldsymbol{\eta}) = (\check{\mathbf{a}} + \boldsymbol{\mathcal{L}}\,\mathbf{z},\;\boldsymbol{\eta})
\end{equation}
where $\boldsymbol{\mathcal{L}}$ is the Cholesky decomposition of the covariance matrix $\boldsymbol{\Psi} = \boldsymbol{\mathcal{L}}\boldsymbol{\mathcal{L}}^\text{T}$. As above, we'll also approximate the covariance of the standardizing transformation with a CURN model, neglecting inter-pulsar correlations for computational efficiency. This amounts to replacing $\boldsymbol{\mathcal{L}}\rightarrow\boldsymbol{\mathcal{L}}_\text{CURN}$ where $\boldsymbol{\mathcal{L}}_\text{CURN}$ is the Cholesky decomposition of the covariance matrix which neglects inter-pulsar correlations, $\boldsymbol{\Psi}_\text{CURN} = \big(\mathbf{F}^\text{T}\tilde{\mathbf{N}}^{-1}\mathbf{F} + \boldsymbol{\varphi}_\text{CURN}^{-1}\big)^{-1} = \boldsymbol{\mathcal{L}}_\text{CURN}\boldsymbol{\mathcal{L}}_\text{CURN}^\text{T}$with $\boldsymbol{\varphi}_\text{CURN}$ being identical to $\boldsymbol{\varphi}$, except inter-pulsar correlations are fixed to zero so the implemented transformation is
\begin{equation}\label{eq:non_gauss_standard_transform_curn}
    (\mathbf{a}, \;\boldsymbol{\eta}) = T^{-1}(\mathbf{z}, \boldsymbol{\eta}) = (\check{\mathbf{a}} + \boldsymbol{\mathcal{L}}_\text{CURN}\,\mathbf{z},\;\boldsymbol{\eta})\,.
\end{equation}

For significantly non-Gaussian priors, such as those with heavy tails, the Laplace approximation is not valid in a large portion of parameter space. Similarly, we rely on inter-pulsar correlations to claim a detection of a stochastic GWB, but the standardizing transformation above does not capture any of these features. This does not bias our inference as the purpose of the standardizing transform is only to approximately de-correlate the parameter space. After the transformation, the posterior which includes inter-pulsar correlations and all non-Gaussian features is evaluated, Eq.~(\ref{eq:non_gauss_post}), to determine the transition probability of the chain. Thus we may infer non-Gaussian features and inter-pulsar correlations while neglecting these contributions in the coordinate transformation. Moreover, the latest analyses \cite{NG15} suggest there is significant evidence for a CURN model under the Gaussian approximation so a coordinate transformation under identical model assumptions will yield an effective reparameterization.

We test our non-Gaussian framework by simulating a stochastic GWB in a collection of simulated pulsars according to Student's t-distribution, which includes non-Gaussian features (for an overview of Student's t-distribution, see e.g. \cite{studentT_moments}). Student's t-distribution is a useful toy example because it generalizes the normal distribution under a simple parameterization, and can achieve heavy non-Gaussian tails. The probability density function for Student's t-distribution is
\begin{equation}\label{eq:student_t_pdf}
    p_\nu(t) = \frac{\Gamma(\frac{\nu + 1}{2})}{\sqrt{\pi\nu}\Gamma(\nu/2)}\bigg(1 + \frac{t^2}{\nu}\bigg)^{-(\nu+1)/2}
\end{equation}
where $\nu$ is the number of degrees of freedom and $\Gamma$ is the gamma function. Student's t-distribution has mean 0 (for $\nu > 1$) and variance $\nu/(\nu-2)$ (for $\nu > 2$). For $\nu=1$ and $\nu\rightarrow\infty$, Student's t-distribution converges to the standard Cauchy and normal distributions, respectively. For finite $\nu$, Student's t-distribution exhibits heavier tails than a Gaussian.

We simulate the timing delays induced by a non-Gaussian stochastic GWB by drawing $2N_fN_p$ independent random variables, $\mathbf{t}$, from Student's t-distribution, Eq.~(\ref{eq:student_t_pdf}), with some chosen (finite) value for $\nu$. Then the prior covariance matrix, $\boldsymbol{\phi}$, for the Fourier coefficients due to an HD-correlated GWB, Eq.~(\ref{eq:prior_cov}), is constructed under a power law spectral model, and we compute the Cholesky decomposition of $\sqrt{(\nu-2)/\nu}\,\boldsymbol{\phi} = \mathbf{L}_\nu\mathbf{L}_\nu^\text{T}$. Finally we color the draws, $\mathbf{t}$, using this Cholesky decomposition $\mathbf{a}_\nu = \mathbf{L}_\nu\mathbf{t}$ and the induced timing delays are $\boldsymbol{\delta t}_\nu = \mathbf{F}\mathbf{a}_\nu$. This simulation determines our prior which exhibits the usual first two statistical moments, $\mathbb{E}[\mathbf{a}_\nu] = \mathbf{0}$ and $\mathbb{E}[\mathbf{a}_\nu\,\mathbf{a}_\nu^\text{T}]=\boldsymbol{\phi}$ as expected for the GWB, but the distribution of Fourier coefficients include higher-order statistical moments due to the initial draws from Student's t-distribution. Because the coloring procedure is a linear coordinate transformation, the prior probability density function consistent with the simulation is derived from the probability density function for the initial independent draws,
\begin{align}\label{eq:PTA_student_t_pdf}
    q(\mathbf{a}_\nu|\boldsymbol{\eta}) &= \tilde{q}(\mathbf{t}|\boldsymbol{\eta})\cdot\text{det}(\partial\mathbf{t}/\partial\mathbf{a}_\nu) \nonumber \\
    &= \bigg[\prod_{i=1}^{2N_fN_p}p_\nu(t_i)\bigg] \cdot\text{det}(\partial\mathbf{t}/\partial\mathbf{a}_\nu)\nonumber \\
    &= \frac{1}{\text{det}(\mathbf{L}_\nu)}\prod_{i=1}^{2N_fN_p}p_\nu((\mathbf{L}_\nu^{-1}\mathbf{a}_\nu)_i)
\end{align}
where we've substituted the inverse transformation $\mathbf{t} = \mathbf{L}_\nu^{-1}\mathbf{a}_\nu$ in the last line. Note that in the $\nu\rightarrow\infty$ limit the GWB simulation procedure and the probability density function Eq.~(\ref{eq:PTA_student_t_pdf}) converge to that of the multivariate normal distribution.

\begin{figure}[H]
    \centering
    \includegraphics[width=0.85\linewidth]{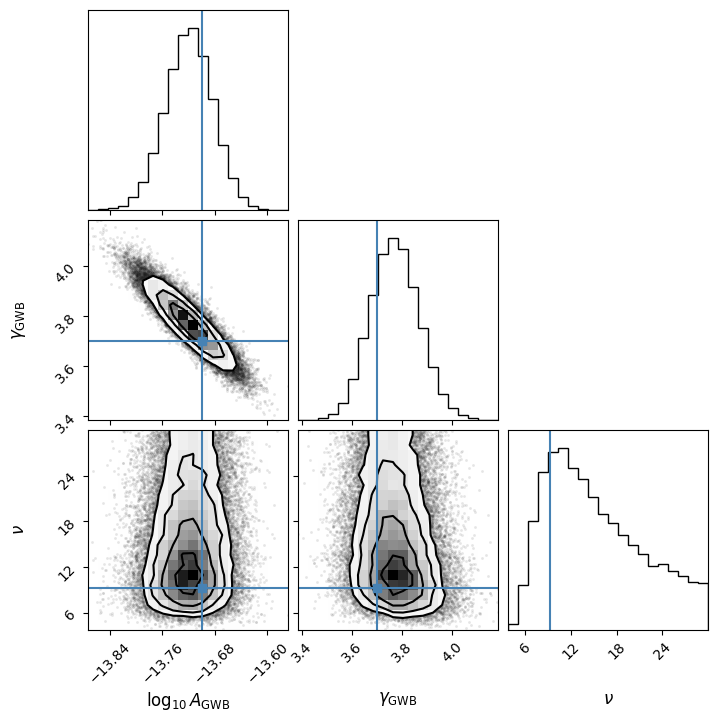}
    \caption{Recovery of a simulated gravitational wave background with non-Gaussian features. The blue lines are injected parameter values. Resolving finite $\nu$ indicates statistical moments beyond Gaussianity.}
    \label{fig:student_t}
\end{figure}

We sample Eq.~(\ref{eq:non_gauss_post}) using Student's t-distribution as a prior, Eq.~(\ref{eq:PTA_student_t_pdf}), under the standardizing transformation Eq.~(\ref{eq:non_gauss_standard_transform_curn}). Data with non-Gaussian features has been simulated such that the mean and covariance of the Fourier coefficients is $\mathbb{E}[\mathbf{a}] = \mathbf{0}$ and $\mathbb{E}[\mathbf{a}\mathbf{a}^\text{T}] = \boldsymbol{\phi}$ respectively, so the standardizing transformation, Eq.~(\ref{eq:non_gauss_standard_transform_curn}) reduces to the usual transformation, Eq.~(\ref{eq:PTA_standard_transform_curn}). In other words, simulating data according to Student's t-distribution induces higher-order statistical moments in the target distribution, but the standardizing transformation, which uses only the estimated first and second moments is unchanged. The recovery of the background parameters is shown in Fig.~(\ref{fig:student_t}). With a sufficiently loud injection, we are able to resolve non-Gaussian features in the background by recovering a finite value for $\nu$.

\bibliographystyle{apsrev4-2}
\bibliography{refs}

\end{document}